\documentclass{aa}%Y
\pdfoutput=1
\usepackage{amsmath}
\usepackage[varg]{txfonts} %Y
\usepackage{natbib}
\usepackage{graphicx}
\usepackage{setspace}%must be 1.5 or double spaced except capts etc-setspace does this
\usepackage{amssymb}
\usepackage{aas_macros}
\usepackage{caption}
\usepackage{subcaption}
\usepackage{subfig}
\usepackage{sidecap}
\usepackage{float}
\usepackage{rotating}
\usepackage{array}
\usepackage{tabularx}
\usepackage{multirow}
\usepackage{subeqn}
\usepackage{pifont}
\usepackage{color}

\bibpunct{(}{)}{;}{a}{}{,}%Y
\begin{document}%Y

\title{The structure and early evolution of massive star forming regions}%Y

\subtitle{Substructure in the infrared dark cloud SDC13}%Y

\author{C. McGuire\inst{1}%Y
	\and G. A. Fuller\inst{1,2}
	\and N. Peretto\inst{3}
	\and Q. Zhang\inst{4}
	\and A. Traficante\inst{1}
	\and A. Avison\inst{1,2}
	\and I. Jimenez-Serra\inst{5}}

%\authorrunning{C. McGuire et al.}			

%\offprints{C. McGuire, \email{catherine.mcguire@postgrad.manchester.ac.uk}}

\institute{%Y
Jodrell Bank Centre for Astrophysics, Alan Turing Building, School of Physics and Astronomy, The University of Manchester, Oxford Road, Manchester, M13 9PL, U.K.
\and UK ALMA Regional Centre node
\and School of Physics and Astronomy, Cardiff University, Queens Buildings, The Parade, Cardiff, CF24 3AA, UK\\ \email{Nicolas.Peretto@astro.cf.ac.uk}
\and Harvard-Smithsonian Center for Astrophysics, 60 Garden Street, Cambridge, MA 02138, USA
\and Department of Physics and Astronomy, UCL, Gower St., London, WC1E 6BT, UK
}

\date{Received 27 July 2015 / Accepted 11 July 2016}%Y

\abstract{Investigations into the substructure of massive star forming regions are essential for understanding the observed relationships between core mass distributions and mass distributions in stellar clusters, differentiating between proposed mechanisms of massive star formation.} 
{We study the substructure in the  two largest fragments (i.e. cores) MM1 and MM2, in the infrared dark cloud complex SDC13. As MM1 appears to be in a later stage of evolution than MM2, comparing their substructure provides an insight in to the early evolution of massive clumps.}
{We report the results of high resolution SMA dust continuum observations towards MM1 and MM2. Combining these data with \textit{Herschel} observations, we carry out RADMC-3D radiative transfer modelling to characterise the observed substructure.} 
{SMA continuum data indicates 4 sub-fragments in the SDC13 region. The nature of the second brightest sub-fragment (B) is uncertain as it does not appear as prominent at the lower MAMBO resolution or at radio wavelengths. Statistical analysis indicates that it is unlikely to be a background source, an AGB star, or the free-free emission of a HII region. It is plausible that B is a runaway object ejected from MM1. MM1, which is actively forming stars, consists of two sub-fragments A and C. This is confirmed by 70$\mu{m}$ \textit{Herschel} data. While MM1 and MM2 appear quite similar in previous low resolution observations, at high resolution, the sub-fragment at the centre of MM2 (D) is much fainter than sub-fragment at the centre of MM1 (A). RADMC-3D models of MM1 and MM2 are able to reproduce these results, modelling MM2 with a steeper density profile and higher mass than is required for MM1. The relatively steep density profile of MM2 depends on a significant temperature decrease in its centre, justified by the lack of star formation in MM2. A final stellar population for MM1 was extrapolated, indicating a star formation efficiency typical of regions of core and cluster formation.} 
{The proximity of MM1 and MM2 suggests they were formed at the similar times, however, despite having a larger mass and steeper density profile, the absence of stars in MM2 indicates that it is in an earlier stage of evolution than MM1. This suggests that the density profiles of such cores become shallower as they start to form stars and that evolutionary timescales are not solely dependent on initial mass. Some studies also indicate that the steep density profile of MM2 makes it more likely to form a single massive central object, highlighting the importance of the initial density profile in determining the fragmentation behaviour in massive star forming regions.}%Y

%Y
\maketitle%Y 

\section{Introduction}
A complete understanding of  the relationship between the distribution of density peaks in regions of high-mass star formation (i.e. the  \emph{core mass function}, CMF) and the distribution of stellar masses in clusters, the \emph{initial mass function}, IMF \citep[e.g.][]{Konyves2010, Alves2007} may provide an insight in to the mechanisms responsible for the formation of massive stars \citep[e.g.][]{Goodwin2007}.

In order to compare the CMF and IMF in a star forming region, we must be able to accurately characterise its substructure i.e. the  level of \emph{fragmentation} it contains. Studying fragmentation in high mass star forming regions requires high resolution observations of the earliest stages of massive star formation \citep[e.g.][]{Beuther2004, Zhang2009, Swift2009, Bontemps2010, Wang2011, Wang2014}. The earliest evolutionary stages of high mass protostars are difficult to identify, but have characteristics consistent with those seen in massive infrared dark clouds (IRDCs). These IRDCs are cold \citep[10--20K; ][]{Pillai2006, Ragan2012}, dense regions within giant molecular clouds \citep[with column densities $>10^{22}$cm$^{-2}$;][]{Peretto2009}, manifesting themselves as regions of extinction against the mid-IR emission from the galactic plane \citep[e.g.][]{Peretto2009}. Here we investigate sub-fragmentation in SDC13, a region comprising 3 Spitzer IRDCs from the \citet{Peretto2009} catalogue (SDC13.174-0.07, SDC13.158-0.073, SDC13.194-0.073) at a distance of $3.6\rm{kpc}$ \citep{Peretto2014}. 

\begin{figure*}
	\centering
        %\begin{subfigure}[t]{0.5\textwidth}
                \includegraphics[width=1\textwidth]{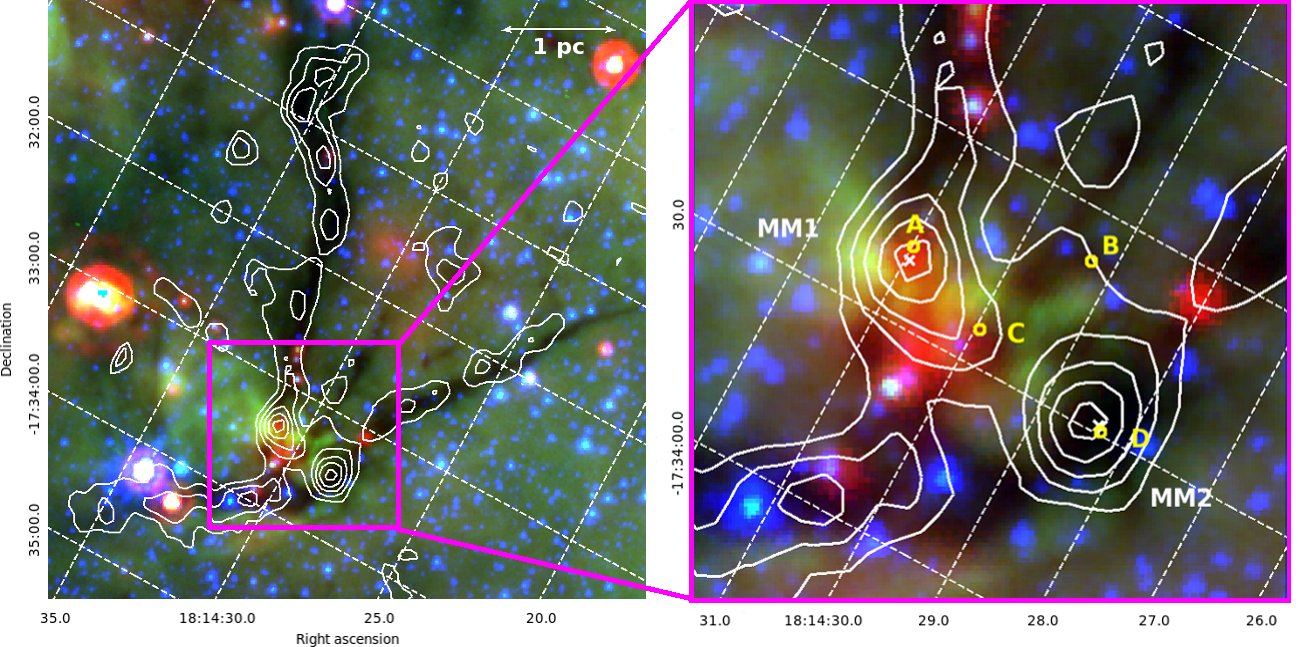}
		%\includegraphics[width=1\textwidth]{modelD_original_FINAL_bold}
                %\caption{Models 1}
        %\end{subfigure}
\caption{A three-colour Spitzer image of SDC13 (24$\rm{\mu{m}}$ in red, 8$\rm{\mu{m}}$ in green and 3.6$\rm{\mu{m}}$ in blue) overlayed with IRAM MAMBO 1.2mm continuum contours (white) at 5, 15, 25, 35 and 45 mJy (left). The highlighted region (right) shows the two largest fragments (i.e. cores) in SDC13, MM1 and MM2, whose positions are marked with white crosses. Four sub-fragments are seen in the SMA 1.3mm continuum observations of the region. The positions of these sub-fragments, A, B, C and D, are marked with yellow circles.}
\label{fig:mambo_spitzer}
\end{figure*}

\citet{Peretto2009} use the term \emph{fragment} to describe local peaks in column density between contours of 8$\rm{\mu}$m opacity for the IRDCs in their catalogue \citep[see Appendix A in][]{Peretto2009}. They find 18 fragments in extinction in SDC13. IRAM 30m MAMBO (Max- Planck-Millimeter-Bolometer) 1.2mm dust continuum observations towards SDC13 indicate fragments of a similar size and position to those seen in extinction, however, the map is dominated by two comparatively large fragments, MM1 and MM2 \citep[see Table \ref{tab:source_properties} below and][]{Peretto2014}. Based on the size of MM1 and MM2 ($\sim$a few times 0.1pc), we consider the term fragment to be analogous to the term core, and will refer to MM1 and MM2 as such. MM1 is not seen in extinction and is associated with 8$\rm{\mu}$m emission, indicative of active star formation. MM2 shows no evidence of star formation activity, lacking both 8$\rm{\mu}$m emission and 24$\rm{\mu}$m emission (associated with warm dust), making it a good candidate for a massive
prestellar core (Figure \ref{fig:mambo_spitzer}).

%\documentclass[12pt,a4paper,final]{report}%size 10 or 12 depending on font used
%\usepackage{setspace}%must be 1.5 or double spaced except capts etc-setspace does this
%\usepackage[left=40mm,right=25mm,top=25mm,bottom=25mm]{geometry}%min left 40mm rest 15mm
%\usepackage{amssymb}
%\usepackage{natbib}
%\usepackage{aas_macros}
%\usepackage{graphicx}
%\usepackage{caption}
%\usepackage{subcaption}
%\usepackage{subfig}
%\usepackage{sidecap}
%\usepackage{float}
%\usepackage{rotating}
%\usepackage{array}
%\usepackage{tabularx}
%\usepackage{multirow}
%\usepackage{subeqn}
%\usepackage{pifont}
%\bibliographystyle{apj3} 
%\doublespacing
%\begin{document}

In this paper we investigate the substructure in MM1 and MM2 using high angular resolution ($<$ 3$^{\prime\prime}$) observations at millimetre wavelengths (Section \ref{sec:obs_and_data}). This is achieved using the Submillimeter Array \citep[SMA\footnote{The Submillimeter Array is a joint project between the Smithsonian Astrophysical Observatory and the Academia Sinica Instituten of Astronomy and Astrophysics and is funded by the Smithsonian Institution and the Academia Sinica}; ][]{Ho2004}, an 8-element radio interferometer located at the summit of  Mauna Kea in Hawaii, to obtain  1.3mm continuum observations of MM1 and MM2 (Section \ref{sec:continuum_image}). The aim is to gain an insight into the mechanisms responsible for massive star formation. As MM2 appears to be in an earlier stage of evolution than MM1, a comparison of the substructure of MM1 and MM2 may  also provide an insight into the early evolution of massive stars and star clusters.

\section{SMA Observations and Data Reduction}\label{sec:obs_and_data}
Observations were performed using 6 antennas of the SMA at 230 GHz. At this frequency the FWHM of the primary beam is $\sim55^{\prime\prime}$.  Both the extended array configuration and compact array configuration were utilised on the 8th March 2012 and the 30th June 2012 respectively. An overview of the observing parameters and the maximum spatial scales that each configuration is sensitive to are given in Table \ref{tab:observing_parameters}. We observed two overlapping fields, centred on MM1 and MM2.

The calibration of the visibility data was performed using MIR, a software package written in IDL for the purpose of reducing SMA data. A time-dependent phase and gain calibration was carried out using quasars 1733-130 and 1743-038 for observations in both the compact and extended configurations. Observations in the lower sideband (LSB) cover the frequency range 216.8-220.8 GHz, and in the upper sideband (USB) 228.8-232.8 GHz. There is a uniform spectral resolution of 0.84 MHz ($\sim$ 1.1 km s$^{-1}$). 

For observations in the extended configuration, the quasar 3c279 was used as the band-pass calibrator; Mars was used to calibrate the flux, and the observed system temperature (T$_{sys}$) varies from 150-200 K. For observations in the compact configuration, Uranus was used to calibrate both the band-pass and the flux, and 120K $<$ T$_{sys}$ $<$ 220K.

After calibration, the visibility data were exported to the radio interferometry data reduction package MIRIAD \citep[Multichannel Image Reconstruction,
Image Analysis and Display,][]{Sault1995} for further processing and image production. Spectral line and continuum data were separated, and line-free continuum data from both antenna configurations in both sidebands were combined for each pointing. In this paper we present results from continuum observations only.

The continuum map shown in Figure \ref{fig:continuum} is a linear mosaic of the pointings towards MM1 and MM2, corrected for primary beam attenuation. The mosaic was produced using the \emph{linmos} command in MIRIAD with the \emph{taper} option selected. This attempts to counteract excessive noise amplification at the edge of the mosaic to produce uniform noise across the whole image \citep[see][]{Sault1996}. The data used combines  observations in the upper and lower side band for both the compact and extended configurations for each pointing. With natural weighting, the map has a synthesised beam with dimensions $3.73^{\prime\prime}\times{2.52}^{\prime\prime}$, position angle (P.A.)$\approx$41.55$^{\circ}$, and a $1\sigma$ rms of $\sim$1 mJy.

\renewcommand{\tabcolsep}{4.7pt}
\begin{table}
\caption{An overview of observing parameters and the largest spatial scales that the SMA is sensitive to in the configurations used.}
% title of Table
\centering
% used for centering table
% is used to refer this table in the text
\begin{tabular}{l l l l l}
% centered columns (4 columns)
\hline\hline
% inserts double horizontal lines
Date			& Config.			& Baseline		& \multicolumn{2}{c}{Max. Spatial Scale}		\\
\cline{4-5}
			& 				& (m)			& (cm)				& (pc)			\\
%units
% inserts single horizontal line
\hline
2012-Mar-08		& Extended			& 44-212		& $3.3\times{10}^{17}$		& 0.1			\\
2012-June-30		& Compact			& 16-77			& $9.0\times{10}^{17}$		& 0.3			\\
% inserting body of the table
\hline
%inserts single line
\end{tabular}
\label{tab:observing_parameters}
\end{table}

\section{SMA 1.3mm Continuum Image}\label{sec:continuum_image}
\begin{figure*}
\centering
\includegraphics[width=0.86\textwidth]{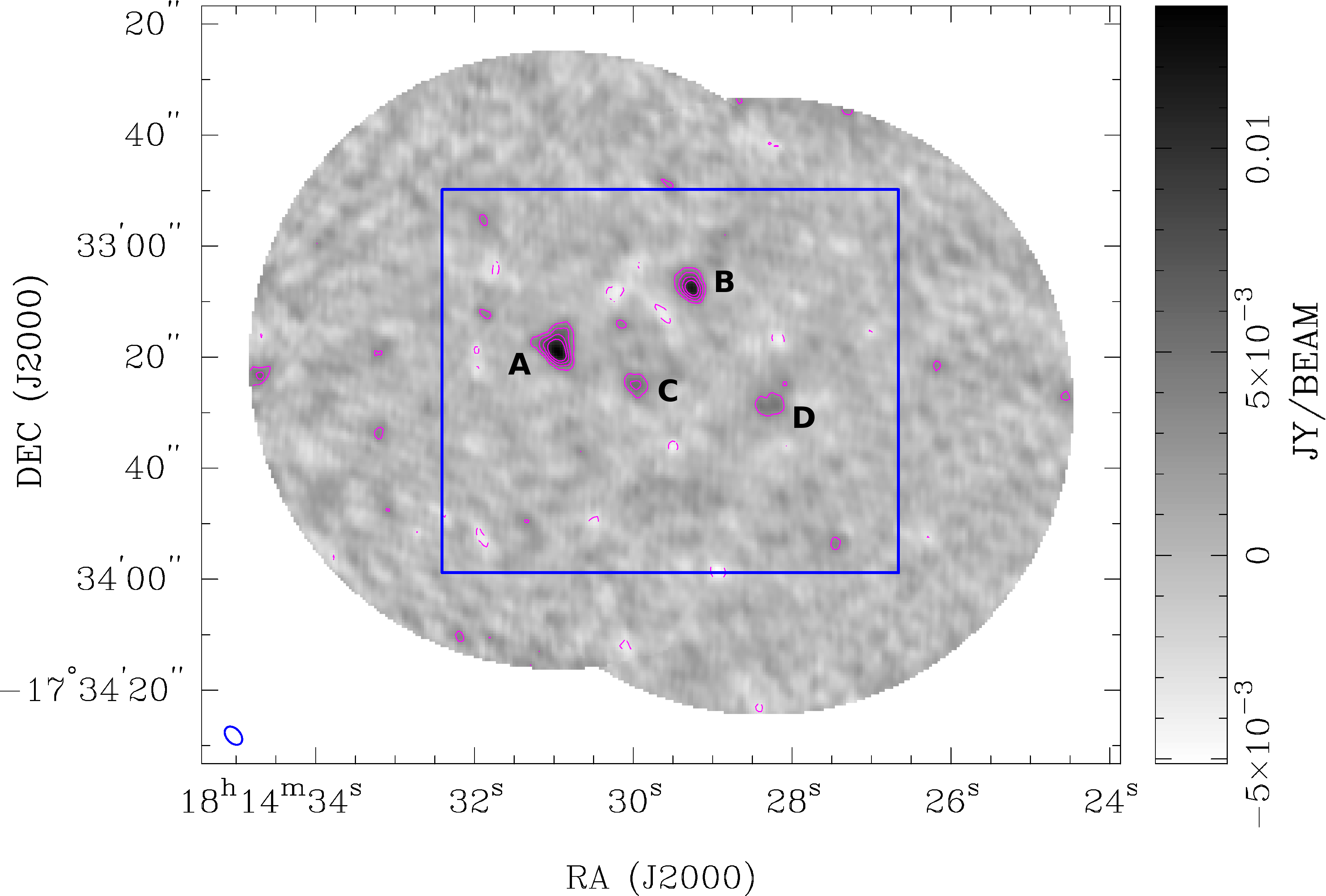}
\caption{The primary beam corrected linear mosaic of the 1.3mm continuum data obtained in two SMA pointings towards MM1 and MM2 (Jy/beam). The central portion, outlined in blue, indicates the region shown in Figure \ref{fig:mambo_contours}. The synthesised beam is shown in the bottom left hand corner of the image. Contours are at $-3\sigma, 3\sigma, 5\sigma, 7\sigma, 9\sigma$ where $\sigma\approx{1}$mJy/beam. Positive contours are indicated by the solid lines and negative contours by the dashed lines. Letters A, B, C and D label the 4 largest sub-fragments seen in the data, defined by the $3\sigma$ contours.}
\label{fig:continuum}
\end{figure*}

The 1.3mm continuum emission towards MM1 and MM2 is shown in Figure \ref{fig:continuum}. We use the term \emph{sub-fragment} to describe the substructures we find within MM1 and MM2 (which is analogous to the term \emph{condensation}, often used to describe the 0.01 pc-scale structures within cores). If we allow sub-fragment boundaries to be defined by $\rm{3\sigma}$ contours, we find that MM1 and MM2 consist of a total of four sub-fragments, which we have labelled A, B, C, D. Table \ref{tab:source_properties} gives the J2000 coordinates and some of the physical properties calculated for A, B, C and D. 

Sub-fragment dimensions a$\times$b of subfragments A and D were estimated based on the $3\sigma$ contours in the 1.3mm continuum map (Figure \ref{fig:continuum}). The position angles (PA) of A and D were determined by fitting each with a 2-D gaussian using the \texttt{Pick Object} function in GAIA (Graphical Astronomy and Image Analysis Tool, part of the \emph{Starlink} astronomical software package). For the weaker subfragments C and D, we estimate diameters to be $4\pm{1}^{\prime\prime}$.

The integrated and maximum flux value for each sub-fragment was determined by performing aperture photometry on the 1.3mm continuum image. This was carried out using the \texttt{imstat} function in MIRIAD. Assuming a typical dust temperature of 15K \citep[e.g. see][]{Rathborne2007, Peretto2014}, the mass for each sub-fragment can then be estimated from its integrated flux using:
\begin{equation}
\rm{M=\frac{F_{\nu}d^{2}}{B_{\nu}(T_{dust})\kappa_{\nu}}},
\label{eq:mdust}
\end{equation}
where $\rm{F_{\nu}=}$ the integrated flux at frequency $\nu$, d is the distance to the source ($\approx{3.6}$kpc) and $\rm{B_{\nu}}$ = the Planck function at a dust temperature $\rm{T_{dust}}$. The opacity at $\nu$ is calculated using
\begin{equation}
\rm{\kappa_{\nu}=10(\nu/1.2 THz)^{\beta}\rm{cm^{2}g^{-1}}} 
\label{eq:knu}
\end{equation}
\citep{Hildebrand1983}, with $\rm{\beta{=1.5}}$ \citep[see][]{Wang2011}, such that $\kappa_{1.3\rm{mm}}=0.8\rm{cm^{2}g^{-1}}$. The masses of MM1 and MM2 given in Table 4 of \citet{Peretto2014} were calculated from the 1.2mm MAMBO integrated flux, assuming a temperature of 15K for MM1 and 12K for MM2, and an opacity $\kappa_{1.2\rm{mm}}=0.5\rm{cm^{2}g^{-1}}$. Thus we apply conversion factors of 0.6 and 0.5 respectively, to obtain the masses for MM1 and MM2 as quoted in Table \ref{tab:source_properties}, allowing a more accurate comparison to the masses we calculate for A, B, C and D.

We obtain masses of $46.8$ M$_{\odot}$ and $40.6$ M$_{\odot}$ for MM1 and MM2 respectively. These are similar to average core masses found in previous studies of massive star forming regions \citep[e.g. ][]{Motte2007, Rathborne2006, Zhang2009}. The masses of sub-fragments A, B, C and D in SDC13  range from $\sim$2--12M$_{\odot}$. These should be considered lower limits as they have not been rescaled to take in to account the filtering of extended emission at 1.3mm \citep[e.g. see][]{Bontemps2010, DuarteCabral2013}, although this is not likely to be a large effect as the sources are close to unresolved. The sub-fragment masses we obtain are in agreement with those obtained by \citet{Wang2011} for sub-fragments in the the star forming region G28.34 (1.4-10.6M$_{\odot}$). 

\begin{figure*}
\centering
\includegraphics[width=1\textwidth]{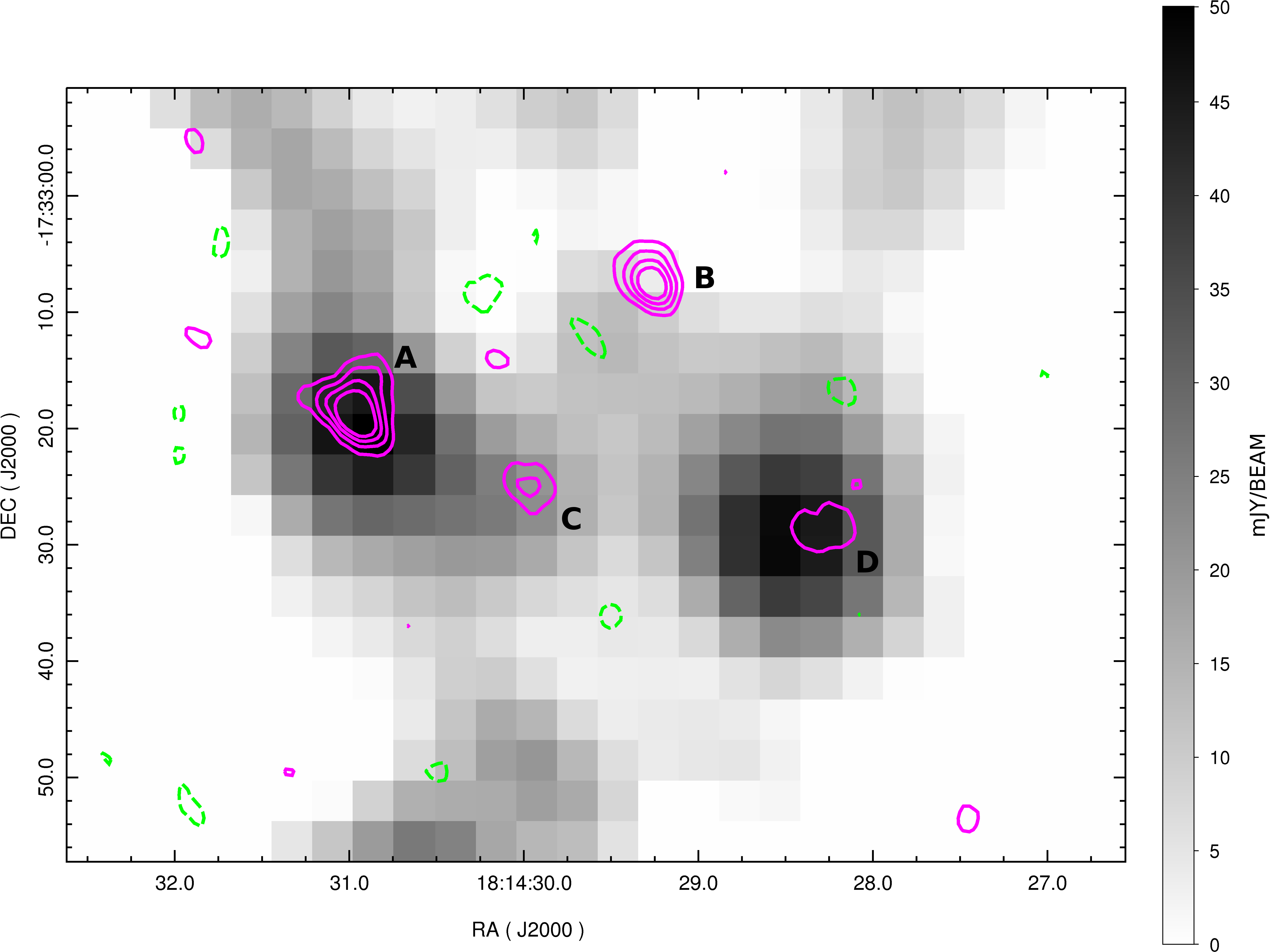}
\caption{SMA 1.3mm continuum data overlayed on 1.2mm MAMBO continuum data (mJy/beam). The contours indicate the SMA continuum at -3, 3, 5,7 and 9 mJy/beam. Positive contours are shown in pink and negative contours are shown dashed in green. Letters A, B, C ,D indicate the 4 \emph{sub-fragments} seen in the region which are identified as peaks in the 1.3mm SMA continuum. The MAMBO beamsize is $10.7^{\prime\prime}$.}
\label{fig:mambo_contours}
\end{figure*}

In Figure \ref{fig:mambo_contours}, the contours from the SMA 1.3mm continuum observations are overlayed on single dish MAMBO 1.2mm continuum data, obtained using the IRAM 30m telescope \citep{Peretto2014}.

Three of the sub-fragments A, C, and D appear to be associated with MM1 and MM2. Sub-fragment B appears to be associated with the slight extension between MM1 and MM2. However, whilst it has a mass approaching that of the brightest sub-fragment A (Table \ref{tab:source_properties}), it is much less prominent in the MAMBO data, indicating that it is not associated with a significant extended envelope of emission. In addition, while it appears that sub-fragment B is close to the position of source MM18 in Figure 1 of \citet{Peretto2014}, the distance between the peak of B and MM18 is larger than its $2^{\prime\prime}$ position uncertainty. Neither does it correspond to any peak in JVLA NH$_{3}$ data (Williams et al., in prep.). In Section \ref{sec:fragmentB} we investigate possible candidates to explain the appearance of sub-fragment B.

Two of the sub-fragments A and C coincide with MM1, with the brightest of the two, sub-fragment A,  at its peak. This could be indicative of fragmentation in MM1. We further investigate this possibility using images from the \textit{Herschel Infrared Galactic Plane Survey} \citep[Hi-GAL,][Section \ref{sec:herschel}]{Molinari2010}.
 
Sub-fragment D coincides with the peak of MM2, however, whereas MM1 and MM2 are of similar size and flux at 1.2mm (Table \ref{tab:source_properties}), sub-fragment D is $\sim$5 times fainter than sub-fragment A at 1.3mm.  An object similar to MM2 is seen in the molecular cloud complex Cygnus X. In their MAMBO survey of Cygnus X, \citet{Motte2007} find a 1.2mm mass of 100 M$_{\odot}$ for the massive cloud core CygX-N40, yet in further studies using the higher resolution of the PdBI \citep{Bontemps2010, DuarteCabral2013} CygX-N40 is barely detectable. This result is attributed to the filtering of extended emission at 1.3mm. The absence of a star in MM2 indicates that it is in an earlier stage of evolution than MM1 which might suggest a more diffuse physical structure, providing a possible explanation for the missing flux at 1.3mm. 

We aim to further understand the physical differences that give rise to such observational differences at higher resolutions. We use RADMC-3D, a software package designed for astrophysical radiative transfer calculations in 1-D, 2-D and 3-D geometries \citep{Dullemond2012} to model MAMBO 1.2mm observations of MM1 and MM2 (Section \ref{sec:modelling}) and use the CASA data reduction software \citep{McMullin2007} to process the resulting images and simulate our SMA 1.3mm observations (Section \ref{sec:simulation}).
%\begin{figure*}[H]
%\centering
%\includegraphics[width=1\textwidth]{images/mambo_contours_2.png}
%\caption{SMA 1.3mm continuum data overlayed on 1.2mm MAMBO continuum data (Jy/beam). The contours indicate the SMA continuum at -3, 3, 5,7 and 9 mJy. Positive contours are shown in pink and negative contours are shown in green. Letters A,B,C,D indicate the 4 \emph{sub-fragments} seen in the region which are identified as peaks in the 1.3mm SMA continuum. The MAMBO beamsize is $10.7^{\prime\prime}$.}
%\label{fig:mambo_contours}
%\end{figure*}

\renewcommand{\tabcolsep}{3.9pt}
\begin{table*}
\caption{Physical properties and J2000 coordinates of MM1 and MM2, and sub-fragments A, B, C and D.}
% title of Table
\small{
%\centering
% used for centering table
% is used to refer this table in the text
\begin{tabular}{l l l l l |l l l l l l l l l l l}
% centered columns (4 columns)
\hline\hline
% inserts double horizontal lines
%		&	&	&		&		&	&	&&					 &		&		&		&		\\						
Frag.		&RA	&Dec	&M$_{1.2mm}$	&R	&Subfrag.	&RA		&Dec		&\multicolumn{2}{c}{Size} 		 &$\rm{R_{eq}}$		&$\rm{R_{d}}$		&PA		&F$_{peak}$	&F$_{int}$	&M$_{1.3mm}$ 	\\
\cline{9-10}
% table heading
		&(h:m:s) &(d:m:s) &(M$_{\odot}$) &(pc)	&		&(h:m:s)	&(d:m:s)	&a$^{\prime\prime}$ & b$^{\prime\prime}$ &(pc) &(pc)			&($^{\circ}$) 	& (mJy)		& (mJy)		& (M$_{\odot}$)	\\
%units
\hline
% inserts single horizontal line
MM1	& 18:14:30.9	& -17:33:20.5	& 	46.8		&	0.26	&	A	& 18:14:31.0	& -17:33:18.5	& 6.6		& 3.3	&	0.04		&0.03	&124.3		&	13.6	&	27.5		&	12.3		\\
	&		&		&			&		&	C	& 18:14:30.0	& -17:33:25.0	& 4.0		& 4.0	&	0.03		&0.02	&...		&	5.6	&	6.7	&	3.0	\\
\hline
MM2	& 18:14:28.3	& -17:33:28.8	&	40.6		&	0.21	&	D	& 18:14:28.3	& -17:33:29.5	& 4.0		& 4.0	&	0.03		&0.02	&...		&	4.9	&	5.6		&	2.5		\\
\hline
	&		&		&			&		&	B	& 18:14:29.3	& -17:33:07.5	& 5.1		& 3.1	&	0.03		&0.02	&38.5		&	12.0	&	15.9		&	7.1		\\
%body of the table
\hline
%inserts single line
\end{tabular}
\tablefoot{M$_{1.2mm}$ is the mass of MM1 and MM2 \citep[extracted from][]{Peretto2014} and multiplied by a conversion factor (see text). R is the deconvolved radius of MM1 and MM2 as calculated by \citet{Peretto2014}. a$\times$b is the non-deconvolved size of A, B, C and D; PA is the position angle, F$_{peak}$ is the peak flux at 1.3mm, F$_{int}$ is the 1.3mm integrated flux, and M$_{1.3mm}$ is the total 1.3mm mass of A, B, C and D. The equivalent radius $\rm{R_{eq}}$=$\rm{\sqrt{A/\pi}}$, where A is the source area. $\rm{R_{d}}$ is the deconvolved equivalent radius. Sub-fragments A and C are associated with MM1, while D is associated with MM2.}
\label{tab:source_properties}
}
\end{table*}

\section{Sub-fragment B}\label{sec:fragmentB}
Sub-fragment B, seen in the SMA data, has a mass (based on its 1.3mm flux) approaching that of sub-fragment A, and yet it is much less prominent in the MAMBO data, indicating that it is not associated with a significant extended envelope of emission. We consider the possibility that sub-fragment B is a background source, an AGB star, the free-free emission of a HII region, or a runaway object.

\citet{Maloney2005} performed a fluctuation analysis on data from the 1.1mm Bolocam Lockman Hole Survey in order to constrain the slope and amplitude of the number count distribution of high redshift galaxies at $\lambda{=}$1.1mm. They find the best-fitting power-law model to have an index $\delta=2.7$, a differential number density at 1 mJy $n_{0}\approx1595$ mJy$^{-1}$deg$^{2}$, and an integrated number density N($>$1 mJy)$\approx{940}$. These three parameters are related by,
\begin{equation}
n_{0}=(\delta{-1})N(>S)S^{\delta{-1}},
\label{eq:probability}
\end{equation}
where S is the peak flux density. The peak flux density of B is S$_{B}\sim$12 mJy (Table \ref{tab:source_properties}). Applying Equation \ref{eq:probability}, we find that N($>$12 mJy)$\approx$13.7. Based on the SMA beamsize, and the overlap of the two pointings, our observations cover an area of $\approx$0.0003 degrees$^2$. The probability of finding a background source within this region is $\sim$0.004. Therefore, it is unlikely that B is background source.

Object B is not listed in any AGB catalogues and, since its angular diameter (see Table \ref{tab:source_properties}) makes it too large to be an AGB star unless it resides within a few parsecs of the Earth \citep[e.g.][]{Villaver2007}, it is  highly unlikely that it is yet to be identified. Using the surface density of dust envelope AGB stars in the solar neighbourhood, calculated by \citet{Olivier2001}, we calculate  the probability of finding an AGB star within the region observed to be $\sim{5.7\times10^{-5}}$.  

In order to determine whether B could be the result of emission from a HII region, we looked at a 5GHz radio continuum image of the SDC13 region obtained by the VLA CORNISH survey. The rms noise in the images is $<$0.4 mJy beam$^{-1}$, which would allow detection of an unresolved UCHII region around a B0 star at 16 kpc \citep{Purcell2013}. Source B is not associated with any CORNISH radio source, so it can not explained by emission from a HII region.

One further possibility is that B is a \emph{runaway} object that has been ejected from MM1, and its continuum emission arises from a circumstellar disk. About 40$\%$ of O stars and $\sim$10$\%$ of B stars are thought to be runaways, and the ejection of stellar embryos from star forming regions, via many body interactions in clustered environments \citep[the dynamical ejection scenario, DES;][]{Poveda1967}, has been proposed as a method of brown dwarf formation \citep[e.g.][]{Bate2009}. Assuming a DES for B, we can estimate its potential ejection velocity based on its distance from the centre of MM1 (i.e. its distance from sub-fragment A) and the approximate  time since ejection. A lower limit on the distance between A and B can be obtained from their observed separation on the sky. This is $\sim${28}$^{\prime\prime}$, equivalent to $\sim1.5\times{10}^{18}$cm at a distance of 3.6kpc.  An upper limit on the time since ejection is equivalent to their formation timescale ($\sim${10}$^{5}$ years for massive prestellar/protostellar objects). Based on these values, we thus calculate a lower limit for the hypothetical ejection velocity of B of $\sim5$kms$^{-1}$. The expected escape velocities for O/B runaways are $\sim$200--400kms$^{-1}$ \citep[ e.g.][]{Gvaramadze2009}. Based on our calculated lower limit for the velocity of sub-fragment B it is plausible that it B is a runaway object.

\section{\textit{Herschel} Hi-GAL Observations}\label{sec:herschel}
Figure \ref{fig:herschel} shows Hi-GAL images at 350, 250, 160 and 70$\mu$m, observed towards SDC13. The $160\mu\rm{m}$ Hi-GAL flux peaks at A, and the $70\mu\rm{m}$ flux peaks over A and C, indicative of star formation activity in this region. Conversely, sub-fragment D is dark in the Hi-GAL $70\mu\rm{m}$ images. This would suggest that D does not contain any embedded sources and may therefore be in an earlier stage of evolution than A and C. 

The SMA continuum data (Figure \ref{fig:continuum}) indicates that MM1 consists of two sub-fragments (A and C). However, A and C appear unresolved in the \textit{Herschel} observations at all wavelengths. At 70$\mu$m, Hi-GAL observations show an elongated source covering the positions of A and C. The resolution of the image is insufficient, however, to confirm whether this elongated source is the result of emission from two objects.
 
Source extraction was performed at all Hi-GAL wavelengths with the \textit{Hyper} (Hybrid Photometry and Extraction Routine) algorithm \citep{Traficante2015}. \textit{Hyper} is an enhanced version of classical aperture photometry, designed to take into account the strong background variability and source crowding typical of Galactic observations. Sources are modelled with 2-D Gaussians, allowing the FWHM to vary between 1 and 2 times the instrumental PSF. The background is subtracted automatically by \textit{Hyper}. It estimates several backgrounds, modelled with polynomials of different orders, and chooses the best background model based on which results in the lowest residuals. The flux is estimated within a 2-D Gaussian region with aperture radii equal to the Gaussians FWHM \citep{Traficante2015}. \textit{Hyper} identified 2 sources at 70$\mu$m, Her1 and Her2. The results of the source extraction and photometry for Her1 and Her2 are shown in Table \ref{tab:herschel_70micron}.

\begin{figure}
	\centering
        \begin{subfigure}[t]{0.5\textwidth}
                \includegraphics[width=1\textwidth]{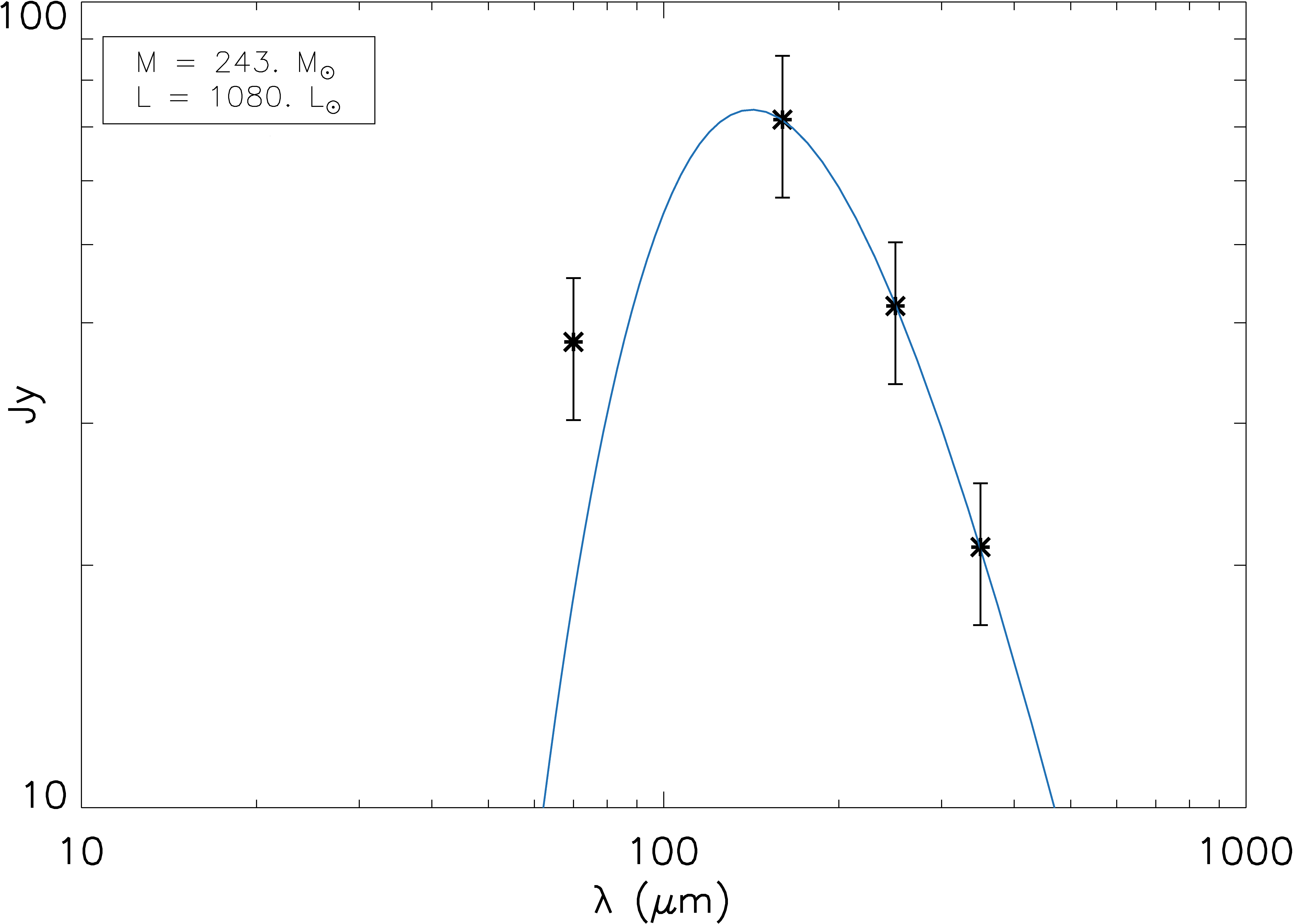}\\
                %\caption{Models 1}
                %\label{fig:models1}
        \end{subfigure}
\caption{SED fit to \textit{Herschel} Hi-GAL data observed towards MM1 at 300, 250, 160, 70$\mu$m, based on a distance of 3.6kpc and using $\beta=1.5$.}
\label{fig:herschel_sed}
\end{figure}

\begin{figure*}
	\centering
        \begin{subfigure}[t]{0.45\textwidth}
                \includegraphics[width=1\textwidth]{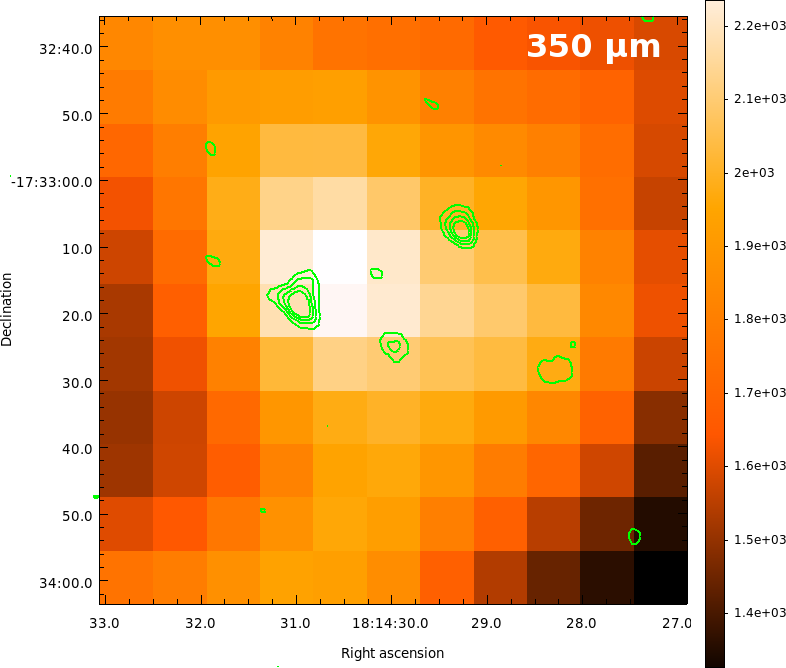}
                %\caption{Models 1}
                %\label{fig:models1}
        \end{subfigure}
        \begin{subfigure}[t]{0.45\textwidth}
                \includegraphics[width=1\textwidth]{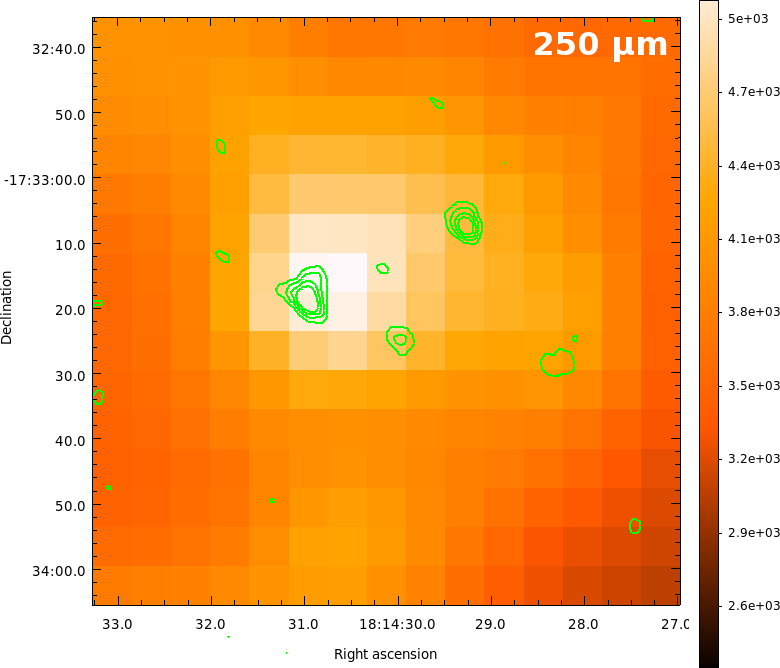}
                %\caption{Models 1}
                %\label{fig:models1}
        \end{subfigure}\\
        \begin{subfigure}[t]{0.453\textwidth}
                \includegraphics[width=1\textwidth]{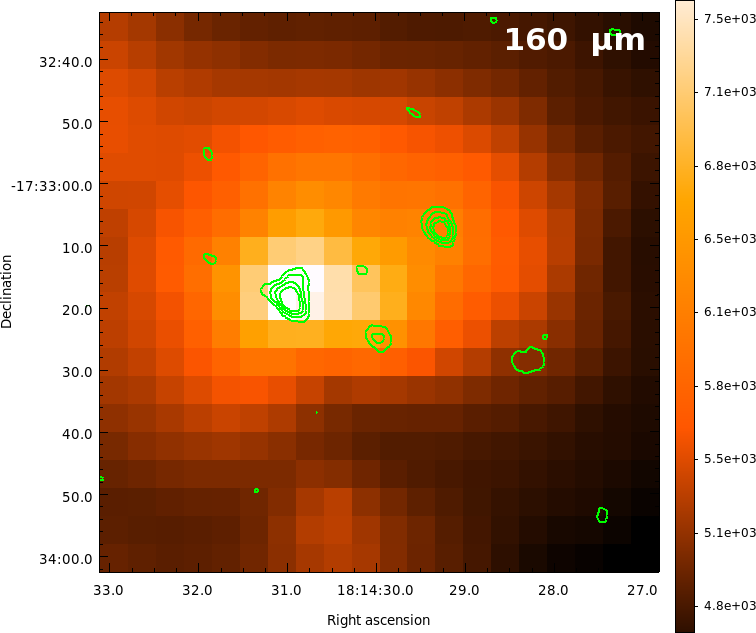}
                %\caption{Models 2}
                %\label{fig:models2}
        \end{subfigure}
        \begin{subfigure}[t]{0.447\textwidth}
                \includegraphics[width=1\textwidth]{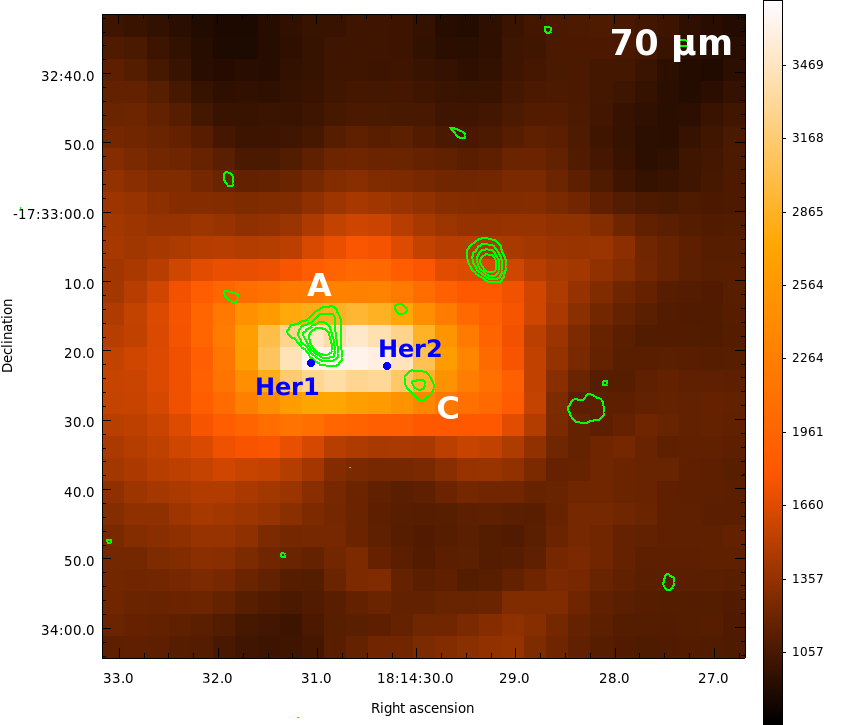}
                %\caption{Models 2}
                %\label{fig:models2}
        \end{subfigure}
\caption{SMA 1.3mm continuum data overlayed on \textit{Herschel} Hi-GAL images (Jy/beam) at 350$\mu$m (top left), 250$\mu$m (top right), 160$\mu$m (bottom left) and 70$\mu$m (bottom right). The contours indicate the SMA 1.3mm continuum at 3, 5, 7 and 9 mJy/beam. Letters A and C indicate 2 of the 4 sub-fragments seen in the region which coincide with MM1. The two sources, Her1 and Her2, indicated in blue, were extracted from the 70$\mu$m \textit{Herschel} data using the \emph{Hyper} algorithm \citep[][see Table \ref{tab:herschel_70micron}]{Traficante2015}.}
\label{fig:herschel}
\end{figure*}

The initial parameters for our RADMC-3D models of MM1 and MM2 (Section \ref{sec:modelling}) are derived based on an SED fit to Hi-GAL observations towards MM1 at 160, 250 and 350 $\mu$m, assuming a distance to SDC13 of d $\simeq3.6$ kpc. We use an elliptical aperture with axes equal to the FWHMs derived from a 2-D Gaussian fit at 250 $\mu$m. The SED fit is performed using a single-temperature greybody model with fixed $\beta$=1.5, and temperature and mass as free parameters. The results of the fit are shown in Figure \ref{fig:herschel_sed}. We find an MM1 luminosity $\rm{L_{Hi-GAL}=1080L_{\odot}}$, radius $\rm{R_{Hi-GAL}=1.93\times10^{18}cm}$ ($\sim$0.6 pc) and core mass $\rm{M_{Hi-GAL}=243M_{\odot}}$ (equivalent to a dust mass $\rm{M_{dust}=2.43 M_{\odot}}$, assuming a gas:dust mass ratio of 100:1). We used the physical properties of MM1 derived from this SED to model MAMBO observations of MM1 and MM2 (Section \ref{sec:modelling}).

\renewcommand{\tabcolsep}{5pt}
\begin{table}
\caption{Photometry results and J2000 coordinates for sources Her1 and Her2.}
% title of Table
\centering
% used for centering table
% is used to refer this table in the text
\begin{tabular}{l l l l l l}
% centered columns (4 columns)
\hline\hline
% inserts double horizontal lines
Name	& RA	& Dec		&PA			& F$_{peak}$	& F$_{int}$		\\
%\cline{4-5}
% table heading
	& (h:m:s)	& (d:m:s)		&($^{\circ}$)		& (Jy)		& (Jy)			\\
%units
\hline
% inserts single horizontal line
Her1& 18:14:31.1	& -17:33:21.6				&156.0		&	0.4	&	14.5$\pm$0.5		\\
Her2& 18:14:30.3	& -17:33:20.5				&152.6		&	0.3	&	13.0$\pm$0.6		\\
% inserting body of the table
\hline
%inserts single line
\end{tabular}
\tablefoot{Data was extracted from 70$\mu{m}$ \textit{Herschel} observations of MM1 using the \emph{Hyper} algorithm \citep{Traficante2015}.}
\label{tab:herschel_70micron}
\end{table}

\section{Modelling with RADMC-3D}\label{sec:modelling}
Using RADMC-3D, we model the dust continuum emission of MM1 and MM2 with 1-D logarithmically spaced, spherically symmetric models with inner radii R$_{in}=1\times10^{16}$cm \citep[based on typical values for best-fit models to high-mass cores given in Table 5 of][]{Williams2005}. This corresponds to $\sim$0.2$^{\prime\prime}$ at the distance of SDC13. Our initial estimates for the outer radius ($\rm{R_{out}=1.93\times10^{18}cm}$) of our model dust clouds and the total dust mass ($\rm{M_{dust}=2.4M_{\odot}}$) are based on the SED fit to the Hi-GAL data for MM1 (Section \ref{sec:herschel}). We use dust opacities calculated by \citet{Ossenkopf1994} for coagulated dust grains with thin ice mantles, at a gas density of $10^5$ cm$^{-3}$ and a \citet{Draine1984} dust grain size distribution. 

Spitzer MIPSGAL observations of the IRDC SDC13 indicate a 24$\mu$m source at the centre of MM1, indicating the presence of a central star (or multiple system). There is no evidence of such a source at the centre of MM2 (Figure \ref{fig:mambo_spitzer}). The effect of including a stellar source at the centre of our models is therefore investigated. We cannot rule out the presence of a multiple system at the centre of MM1, however for simplicity we use a single object at the centre of our models. We base the luminosity of the central star ($\rm{L_{\star}=1080L_{\odot}}$) on the luminosity of MM1, which is derived from the Hi-GAL data (Section \ref{sec:herschel}). We estimate the contribution of the dust emission to the total luminosity (in the absence of a star) using a modified blackbody function \citep[e.g. see][]{Battersby2011} at a temperature of 15 K, normalised to the MM1 1.2mm MAMBO flux. We find a contribution of $\sim$14L$_{\odot}$ from dust emission, which is negligible compared to the total luminosity. The remaining stellar properties (mass $\rm{M_{\star}=7.36M_{\odot}}$, radius $\rm{R_{\star}=4.94R_{\odot}}$ and temperature $\rm{T_{\star}=14914}$ K, corresponding to a class B ZAMS star) are calculated using typical relationships between the luminosity, mass, radius and temperature of main sequence stars \citep{Schulz2012}. 

For prestellar cores (without a central source), the dominant source of radiation will be external. To account for this, we include an external radiation field in our models based on the ISRF in the solar neighbourhood. Our models cover wavelengths $0.01\leq{\lambda}\leq 1000 \rm{\mu{m}}$, therefore our model ISRF covers emission in this range, incorporating contributions from the cosmic microwave background (CMB), infrared emission from dust, and photons of starlight \citep{Draine2011}.

The earliest models of core collapse describe isolated spherical cores which are pressure bounded, isothermal and non-fragmenting i.e. \emph{Bonnor-Ebert} (BE) spheres \citep{Bonnor1956, Ebert1955}. For a critical BE sphere on the verge of gravitational collapse, there are a family of solutions corresponding to spheres with different initial density distributions. Each consist of a uniform density central core (whose size and density varies with temperature) and an outer envelope with density $\rho\propto{r}^{-2}$ (where r is the radius). At one extreme of this group of models is the  \citet{Shu1977} solution, which describes the \emph{inside-out} collapse of a singular isothermal sphere (SIS) i.e. a sphere with infinite central density and $\rho\propto{r}^{-2}$. At the other extreme of the family of BE spheres, the Larson-Penston solution \citep{Larson1969, Penston1969} describes the collapse of a cloud with uniform density, that acquires large infall velocities at all radii.

During the inside-out collapse of a SIS, described by \citet{Shu1987}, the density increases fastest in central regions, with collapse  occurring first at the centre and then propagating outwards via an expansion wave \citep{Shu1977}. The result of the collapse is an infalling central region in free-fall collapse ($\rho\propto{r}^{-1.5}$) surrounded by a static envelope (retaining the isothermal density distribution ($\rho\propto{r^{-2}}$).  Other theoretical models invoke different density profiles at various points during the process of collapse \citep[e.g.][]{Foster1993, Basu1994, Whitworth2001}, however, \citet{Foster1993} found that following the formation of the central protostar, density profiles tend towards those seen for SISs. More recently, based on numerical simulations of core and protostar formation in supersonic flows, \citet{Gong2009, Gong2011} find  the $\rho\propto{r}^{-2}$ density profile to be  an \emph{attractor} for the collapse of a molecular cloud core at the point of protostar formation, regardless of the mechanism responsible for the collapse. Similar results have been found for models of massive cores \citep{Tilley2004}.

Observations of low mass star forming regions \citep[e.g.][]{Young2003} show core density profiles $\rho\propto{r}^{\alpha}$ with $-2.0\leq\alpha\leq{-1.5}$, in agreement with the predictions of models of collapsing cores \citep[e.g.][]{Shu1987, Foster1993}. However the density distributions of high mass cores have been more difficult to characterise due to the relative scarcity of massive cores and their more complex environments. \citet{Beuther2002} presented a study on a large sample of massive star forming regions and found a typical value of $\alpha=-1.6$ for their density distributions. Additional previous modelling and observations of high mass star forming regions have found a range of density distributions with ${-2.25}\leq{\alpha}\leq{0}$ \citep[e.g.][]{Wolfire1994, Garay1990, Hatchell2003, Williams2005}. As such, we model our spherical dust clouds with density distributions $\rho\propto\rm{r}^{\alpha}$ (where ${-3.0}<{\alpha}\leq{0}$). The upper limit on $\alpha$ is a consequence of our mass normalization (see Equation \ref{eq:dust_mass}).

\begin{figure*}
	\centering
        \begin{subfigure}[t]{0.4\textwidth}
                \includegraphics[width=\textwidth]{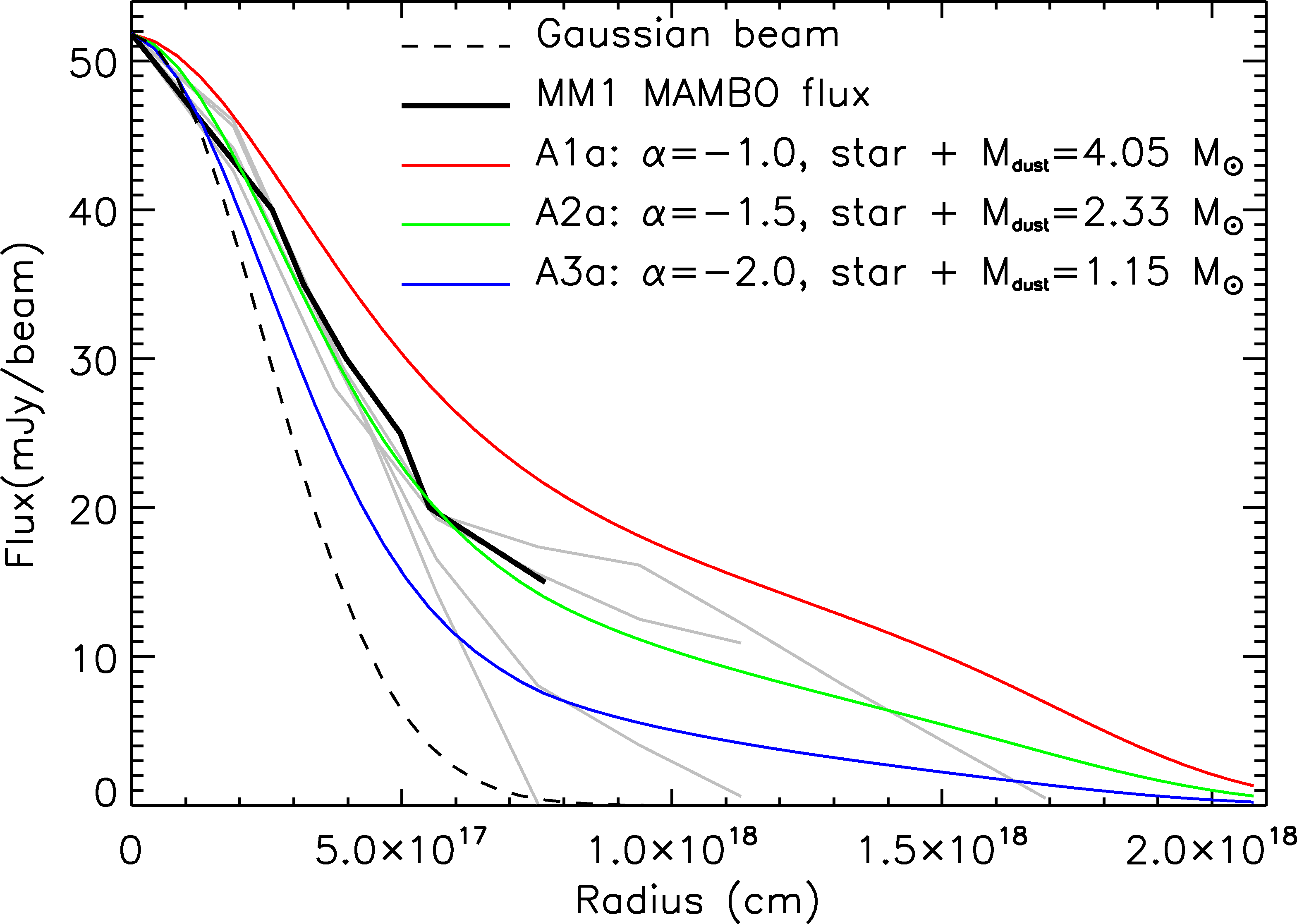}
                \label{subfig:modelAandD_slice_A}
        \end{subfigure}
        \begin{subfigure}[t]{0.4\textwidth}
                \includegraphics[width=\textwidth]{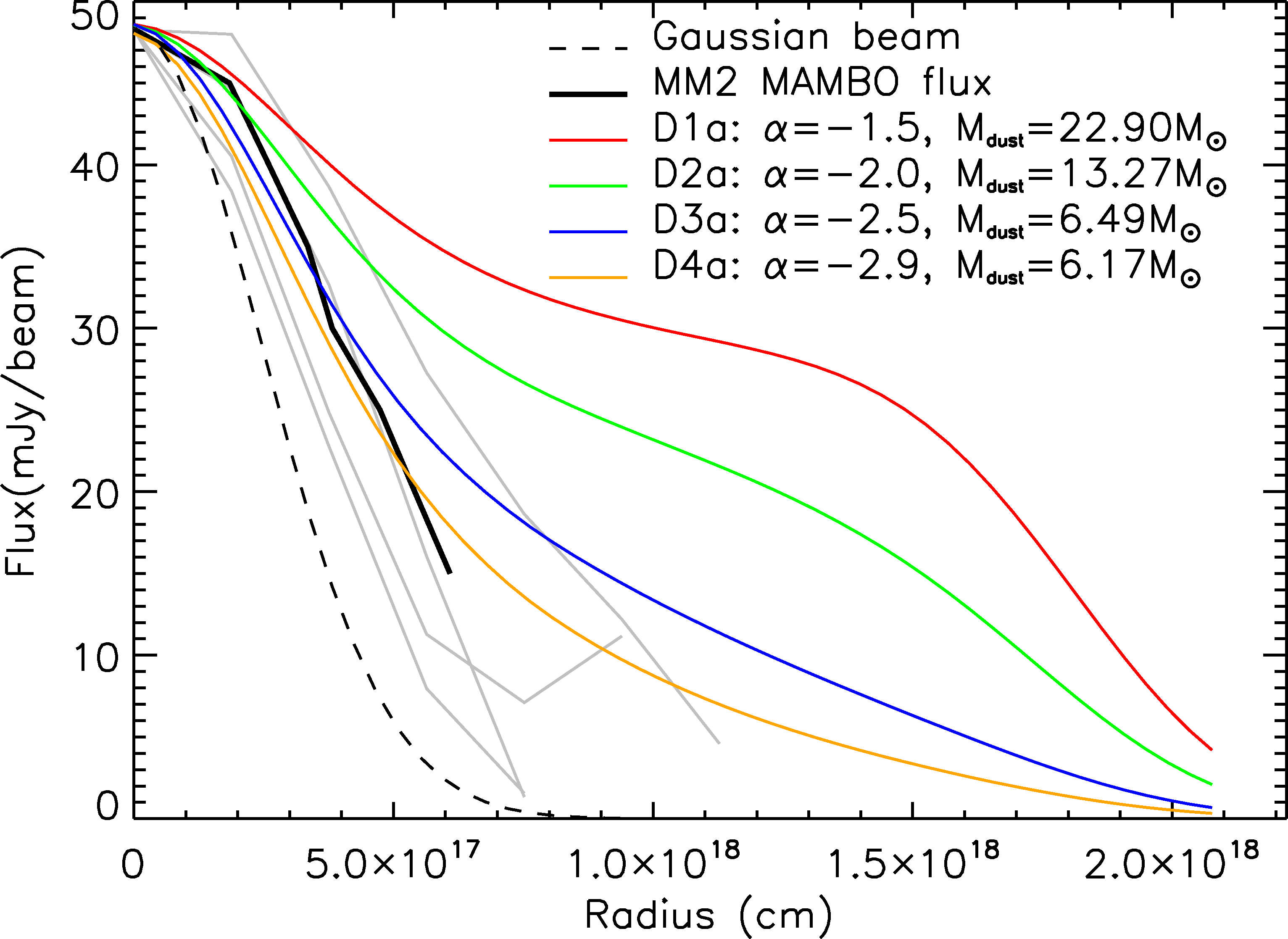}
                \label{subfig:modelAandD_slice_logA}
        \end{subfigure}
	\vspace{5pt}
        \begin{subfigure}[t]{0.4\textwidth}
                \includegraphics[width=\textwidth]{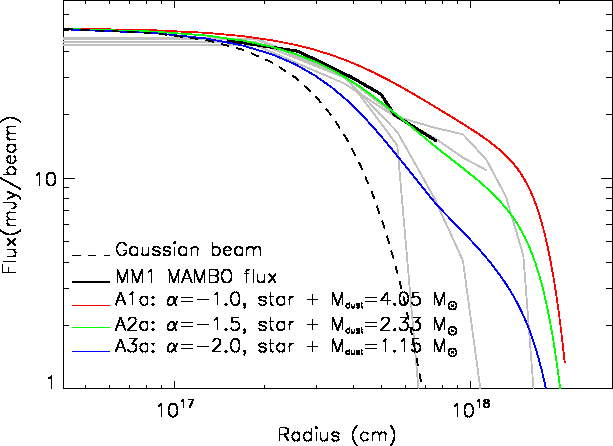}\\
                \label{subfig:modelAandD_slice_D}
        \end{subfigure}
        \begin{subfigure}[t]{0.4\textwidth}
                \includegraphics[width=\textwidth]{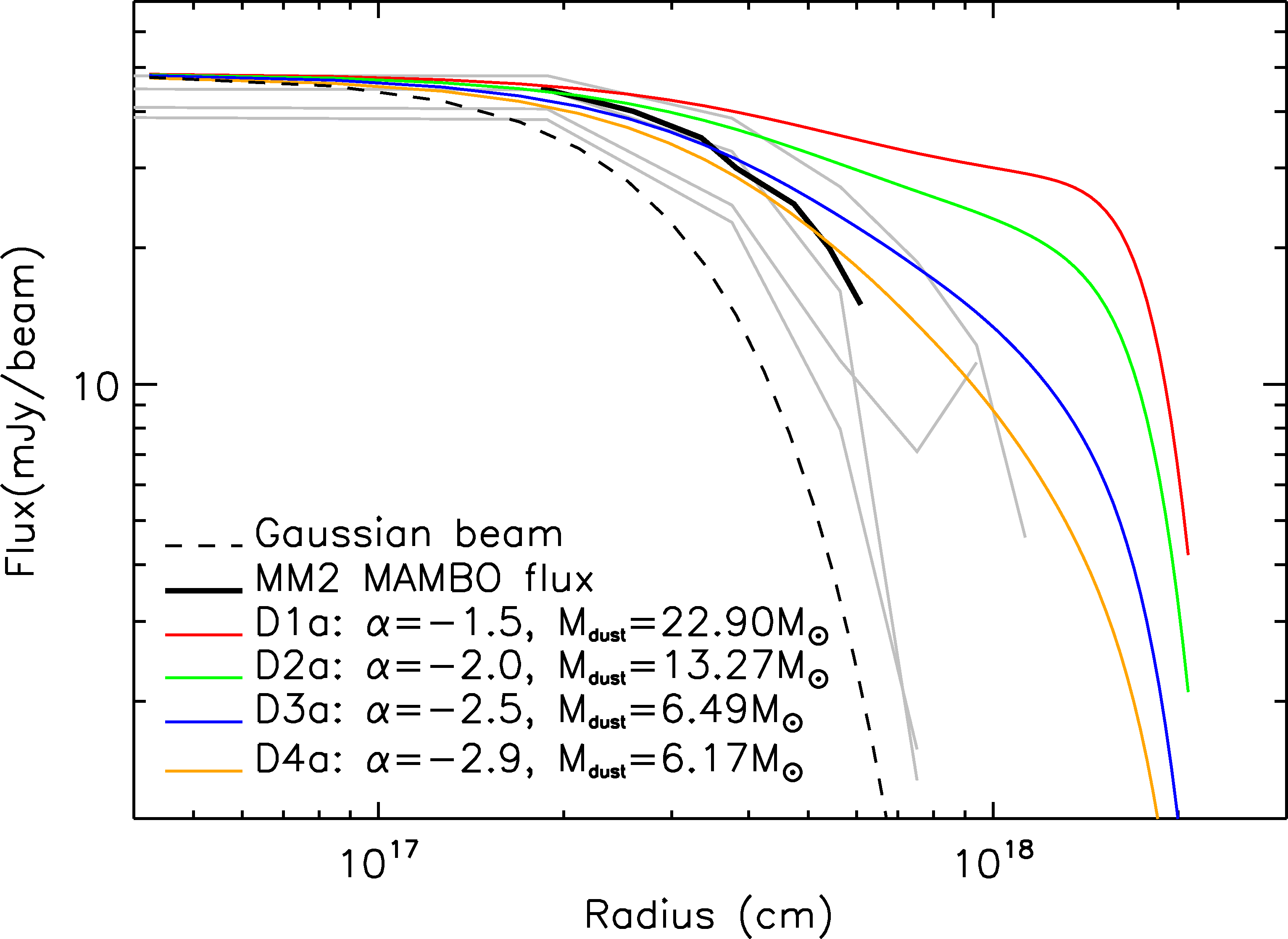}\\
                \label{subfig:modelAandD_slice_logD}
        \end{subfigure}
\caption{Comparison of the observed MAMBO 1.2mm flux distributions of MM1 (left) and MM2 (bottom) with RADMC-3D model fluxes, obtained using various model parameters. The solid black lines indicate the observed fluxes plotted against the equivalent radius R$\rm{_{eq}}=\sqrt{{\rm{A_{n\sigma}}/{\pi}}}$ (where $\rm{A_{n\sigma}}$ is the area contined within contours $\rm{n\sigma}$, for $\rm{n}=$3, 4,  6, 7, 8, 9 and $\sigma\sim5$mJy/beam). The grey lines indicate perpendicular slices of observed flux against radius; these slices give an indication of the range of variation in the observed flux distributions in MM1 and MM2. The coloured lines show the model fluxes against radius for RADMC-3D models with various dust distributions, as indicated in the key. The model dust masses $\rm{M_{dust}}$ required to match each peak model flux with the peak observed flux are also given. Models are labelled A1a to A3a (for those fit to MM1 data) and D1a to D4a (for those fit to MM2 data). The dashed curves indicate the MAMBO beam ($10.7^{\prime\prime}$).}
\label{fig:modelAandD_slice}
\end{figure*}

Figure \ref{fig:modelAandD_slice} compares the MAMBO 1.2mm continuum flux distributions observed towards MM1 and MM2 for a range of power law density distributions. Using the initial Hi-GAL derived estimate for the core radius $\rm{R_{out}=R_{Hi-GAL}}=1.93\times10^{18}\rm{cm}$,  we obtain good fit to the slope of the MAMBO 1.2mm observations of MM1 using a model with a central source and a density distribution $\rho\propto{r}^{-1.5}$ (model A2a). This is similar to the observed density profiles of protostellar cores in low-mass star forming regions, and is suggestive of free-fall collapse \citep{Shu1977, Shu1987}, consistent with infall around the central star seen in the Spitzer data (Figure \ref{fig:mambo_spitzer}). In order to also fit the peak of the MM1 observations using this density distribution, we require a model dust mass (i.e. the mass of dust
only) $\rm{M_{dust}=2.33M_{\odot}}$. 

\begin{figure*}
	\centering
        \begin{subfigure}[t]{0.4\textwidth}
                \includegraphics[width=\textwidth]{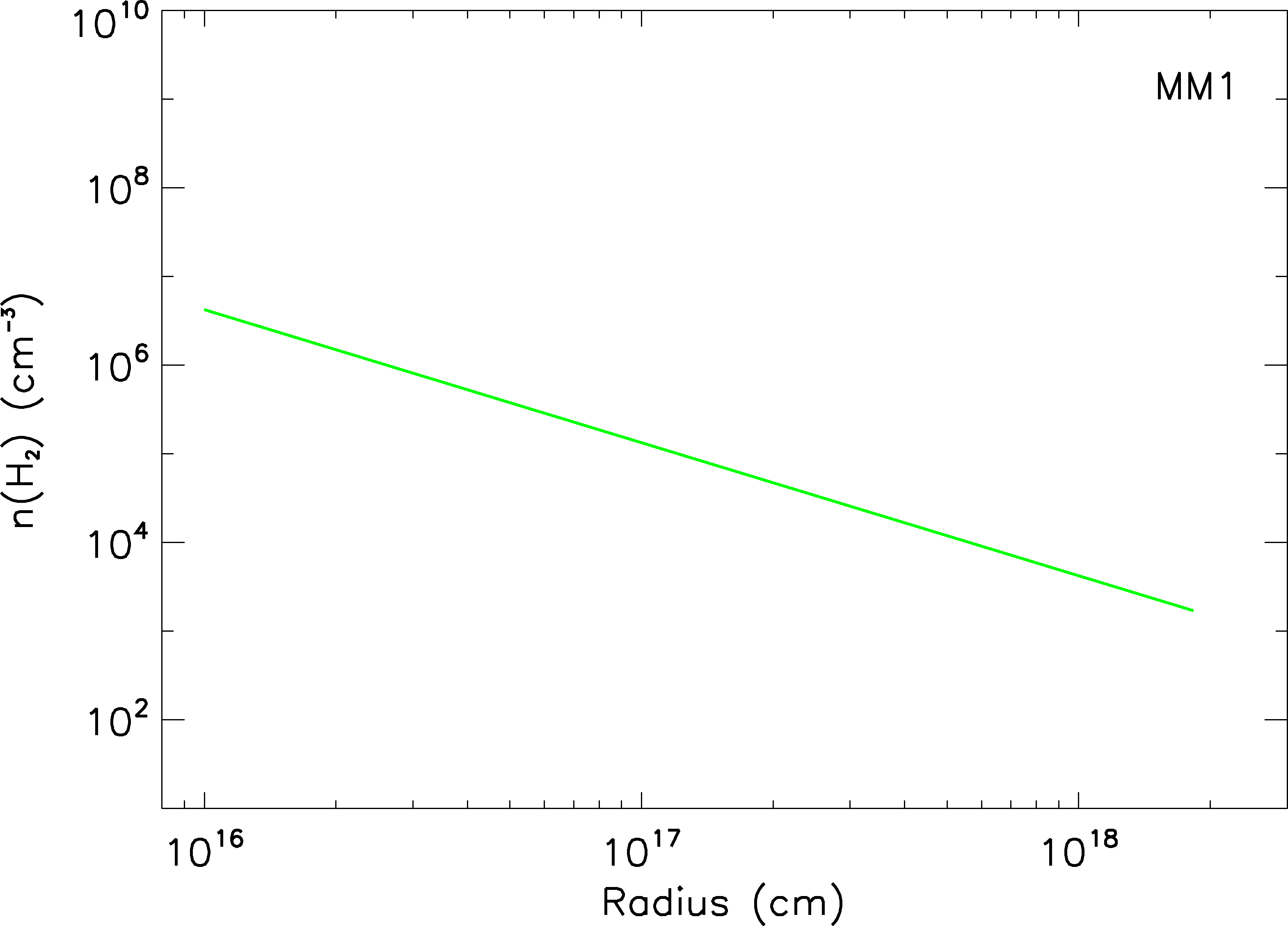}
                \label{subfig:densityvsradA}
        \end{subfigure}
        \begin{subfigure}[t]{0.4\textwidth}
                \includegraphics[width=\textwidth]{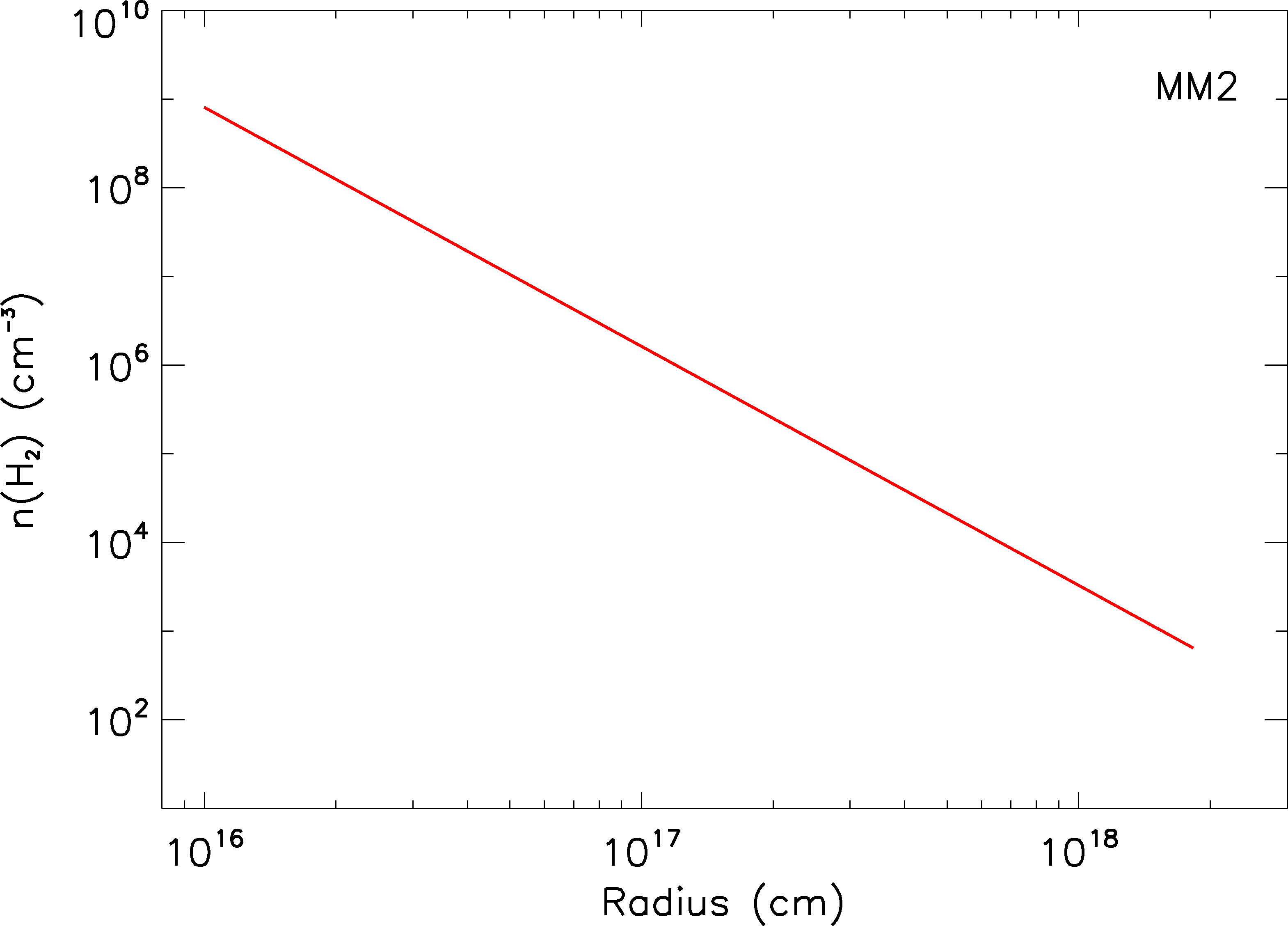}
                \label{subfig:densityvsradD}
        \end{subfigure}
	\vspace{5pt}
        \begin{subfigure}[t]{0.4\textwidth}
                \includegraphics[width=\textwidth]{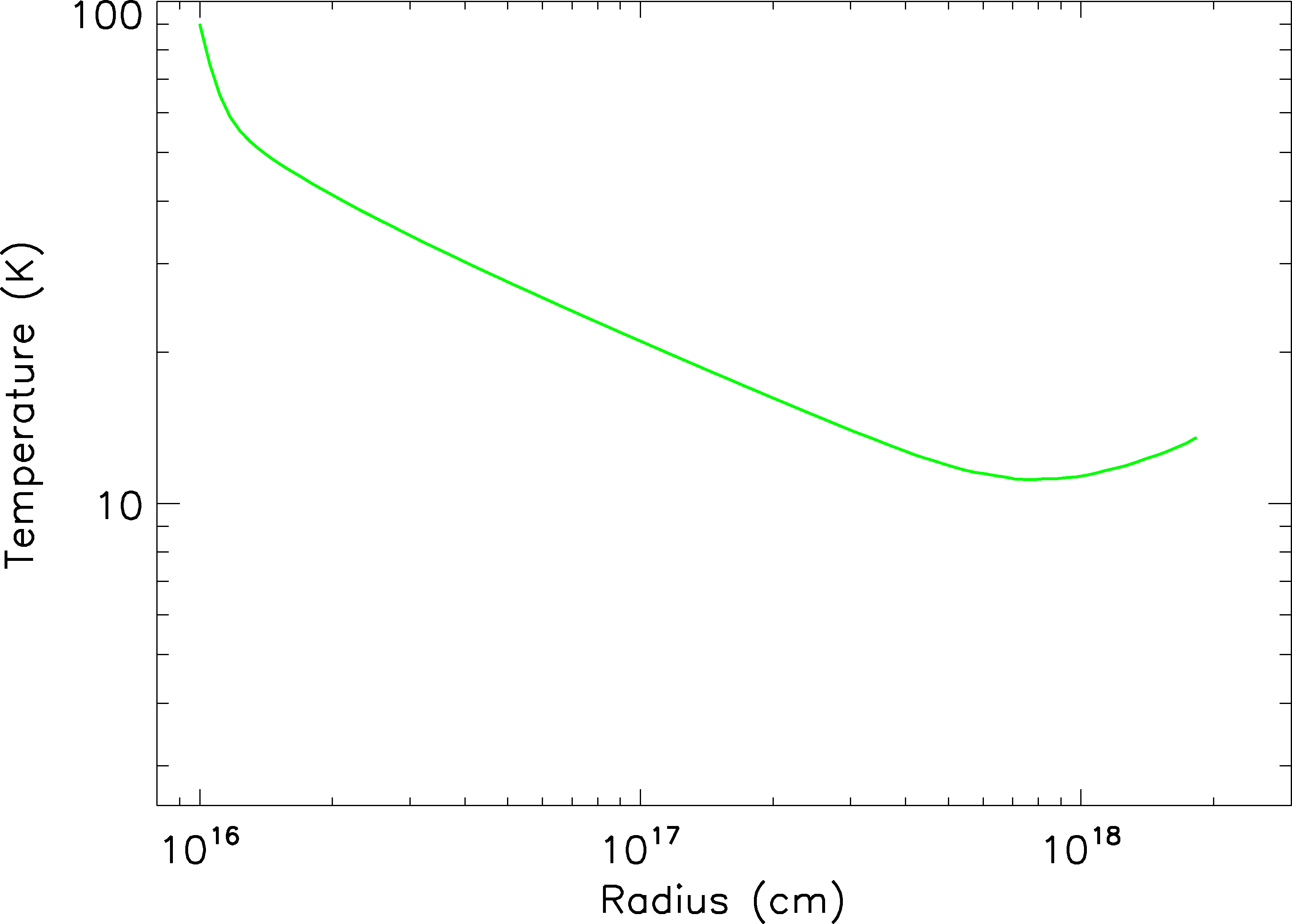}\\
                \label{subfig:tempvsradA}
        \end{subfigure}
        \begin{subfigure}[t]{0.4\textwidth}
                \includegraphics[width=\textwidth]{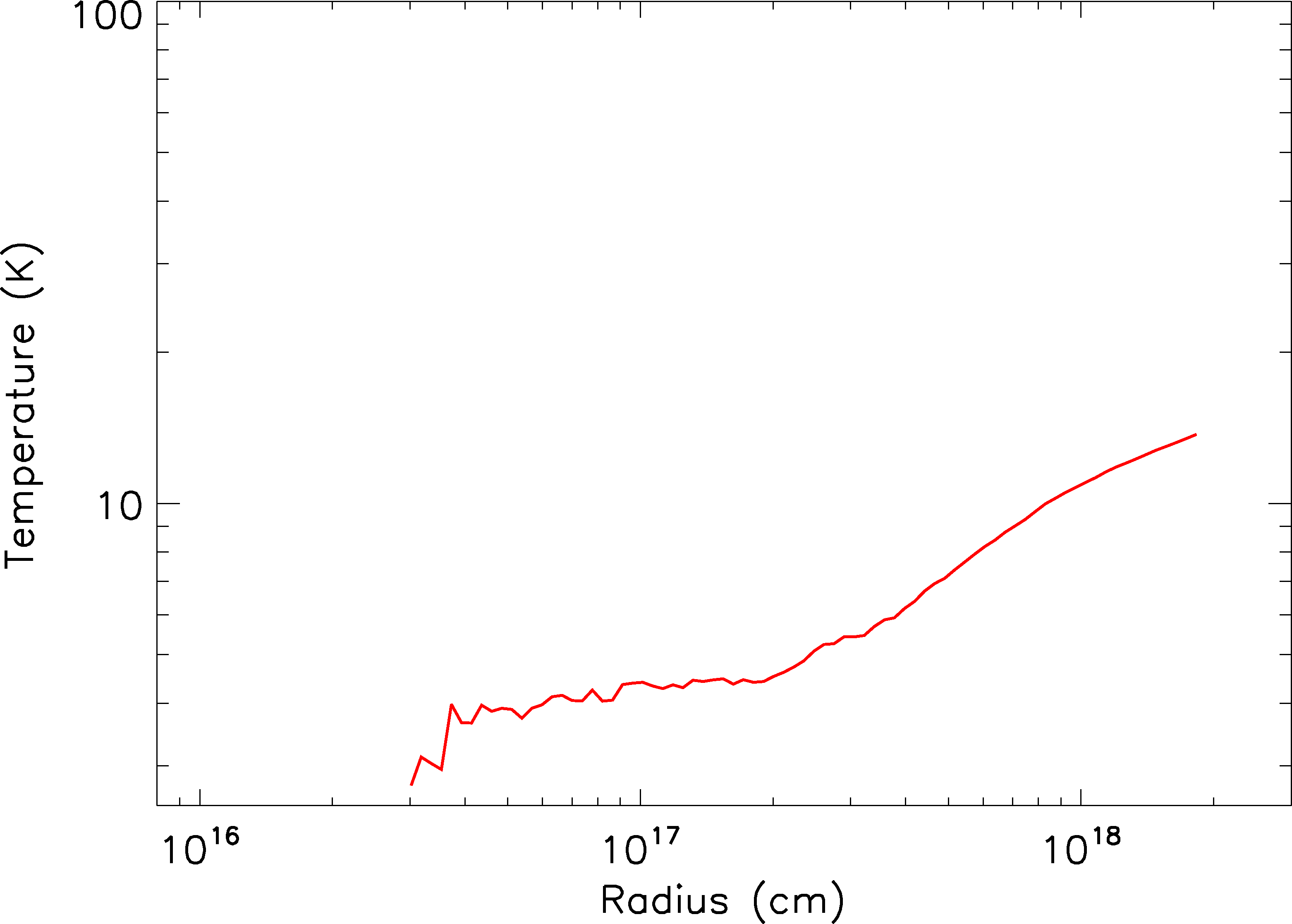}\\
                \label{subfig:tempvsradD}
        \end{subfigure}
\caption{Density and temperture profiles of the best fit models (with $\rm{R_{out}=R_{Hi-GAL}}$) to MAMBO observations of MM1 (left) and MM2 (right; models A2a and D3a respectively).}
\label{fig:density_temp}
\end{figure*}

We initially modelled MM2 as a dust envelope without a central source, with $\rm{R_{out}=R_{Hi-GAL}}$. A reasonable fit to the slope of the MM2 MAMBO data is then obtained for a model with a power law density distribution $\rho\propto{r}^{\alpha}$, with $-2.9<{\alpha}<-2.5$ (models D3a and D4a respectively). This is steeper than expected for an SIS collapse scenario. The best fit models to MM2 have dust masses in the range $\rm{{6.17M_{\odot}}}<{\rm{M_{dust}}}\leq\rm{{6.49M_{\odot}}}$. Inspection of the MIPSGAL images suggests there may be some heating due to an enhanced radiation field in this region (Figure \ref{fig:mambo_spitzer}). In principle this would lower the mass of MM2, but the effect would be subtle. This heating may however account for the spread in the radial profile of MM2 (e.g. see grey lines in Figure \ref{fig:modelAandD_slice}). The temperature and density profiles of the best fit models (with $\rm{R_{out}=R_{Hi-GAL}}$) to MAMBO observations of MM1 and MM2  are shown in Figure \ref{fig:density_temp}. This figure shows that the temperature profile of the best fit model to MM2 drops down to $\sim4-5$ K at its centre. While temperatures as low as 6 K have been seen in some sources \citep[e.g.][]{Harju2008}, massive star forming regions and IRDCs are typically observed to have temperatures of $\sim{10-15}$ K. We therefore investigated the effect of setting a lower temperature profile limit of 10 K on the resulting RADMC-3D model density profiles. This temperature limit could be justified by the effect of cosmic ray heating, for example. We found than while the 10 K lower temperature limit did result in a shallower density distribution for the best fit model to MM2 ($\alpha\sim{2.0}$) this is still steeper than the density profile found for MM1. This fixed minimum temperature did however result in a decreased model mass of $\sim\rm{{230 M_{\odot}}}$ for MM2.  

In an attempt to further improve the fit to the MAMBO observations of MM2 we tried and successfully fitted a number of different models to the MM2 data which correspond to a variety of physical conditions. These included models with truncated density distributions \citep[expected for cold cores embedded in warmer gas;][]{Fischera2014}; and two-part power law density profiles, consistent with observations of central flattening in prestellar cores \citep[e.g.][]{Beuther2002,Andre2000}. The SMA data provides additional constraints on these models. In Section \ref{sec:simulation}, we describe the use of the CASA data reduction software \citep{McMullin2007} to process our model images and produce SMA 1.3mm simulations to investigate if we are able to reproduce our 1.3mm observations (Figure \ref{fig:continuum}). 
%\documentclass[12pt,a4paper,final]{report}%size 10 or 12 depending on font used
%\usepackage{setspace}%must be 1.5 or double spaced except capts etc-setspace does this
%\usepackage[left=40mm,right=25mm,top=25mm,bottom=25mm]{geometry}%min left 40mm rest 15mm
%\usepackage{amssymb}
%\usepackage{natbib}
%\usepackage{aas_macros}
%\usepackage{graphicx}
%\usepackage{caption}
%\usepackage{subcaption}
%\usepackage{subfig}
%\usepackage{sidecap}
%\usepackage{float}
%\usepackage{rotating}
%\usepackage{array}
%\usepackage{tabularx}
%\usepackage{multirow}
%\usepackage{subeqn}
%\usepackage{pifont}
%\bibliographystyle{apj3} 
%\doublespacing
%\begin{document}

%\documentclass[12pt,a4paper,final]{report}%size 10 or 12 depending on font used
%\usepackage{setspace}%must be 1.5 or double spaced except capts etc-setspace does this
%\usepackage[left=40mm,right=25mm,top=25mm,bottom=25mm]{geometry}%min left 40mm rest 15mm
%\usepackage{amssymb}
%\usepackage{natbib}
%\usepackage{aas_macros}
%\usepackage{graphicx}
%\usepackage{caption}
%\usepackage{subcaption}
%\usepackage{subfig}
%\usepackage{sidecap}
%\usepackage{float}
%\usepackage{rotating}
%\usepackage{array}
%\usepackage{tabularx}
%\usepackage{multirow}
%\usepackage{subeqn}
%\usepackage{pifont}
%\bibliographystyle{apj3} 
%\doublespacing
%\begin{document}

%(see Table \ref{tab:twopart_bestfit})
%\documentclass[12pt,a4paper,final]{report}%size 10 or 12 depending on font used
%\usepackage{setspace}%must be 1.5 or double spaced except capts etc-setspace does this
%\usepackage[left=40mm,right=25mm,top=25mm,bottom=25mm]{geometry}%min left 40mm rest 15mm
%\usepackage{amssymb}
%\usepackage{natbib}
%\usepackage{aas_macros}
%\usepackage{graphicx}
%\usepackage{caption}
%\usepackage{subcaption}
%\usepackage{subfig}
%\usepackage{sidecap}
%\usepackage{float}
%\usepackage{rotating}
%\usepackage{array}
%\usepackage{tabularx}
%\usepackage{multirow}
%\usepackage{subeqn}
%\usepackage{pifont}
%\bibliographystyle{apj3} 
%\doublespacing
%\begin{document}
%\end{document}

%\input{./fig_all_truncated_v2.tex}
%\input{./tab_twopart_bestfit_v2.tex}
%\documentclass[12pt,a4paper,final]{report}%size 10 or 12 depending on font used
%\usepackage{setspace}%must be 1.5 or double spaced except capts etc-setspace does this
%\usepackage[left=40mm,right=25mm,top=25mm,bottom=25mm]{geometry}%min left 40mm rest 15mm
%\usepackage{amssymb}
%\usepackage{natbib}
%\usepackage{aas_macros}
%\usepackage{graphicx}
%\usepackage{caption}
%\usepackage{subcaption}
%\usepackage{subfig}
%\usepackage{sidecap}
%\usepackage{float}
%\usepackage{rotating}
%\usepackage{array}
%\usepackage{tabularx}
%\usepackage{multirow}
%\usepackage{subeqn}
%\usepackage{pifont}
%\bibliographystyle{apj3} 
%\doublespacing
%\begin{document}
%\end{document}

\section{CASA Simulation of SMA Observations}\label{sec:simulation}
Initially, the RADMC-3D models of MM1 and MM2 were added linearly into a blank field to create a single sky model using the CASA imaging toolkit functions. The CASA task \texttt{simobserve} was then used to generate visibility data. This created four measurement sets, one for each SMA array configuration at each sideband. The date and hourangle were set such that the UTC and date for each observation matched the true observations. The zenith opacity was set to give T$_{sys}$ values for the simulations which matched the highest values measured in the SMA observations with each array. The on source time, comprising multiple scans toward two pointings, matched the observed data in length. An artificial phase calibration cycle was included such that the \emph{time coverage} of the uv-plane was as close as possible to real observations. The measurement sets (MS) for each sideband for each array were then concatenated together, giving a single MS. 

\begin{figure*}
	\centering
        \begin{subfigure}[t]{0.56\textwidth}
                \includegraphics[width=1\textwidth]{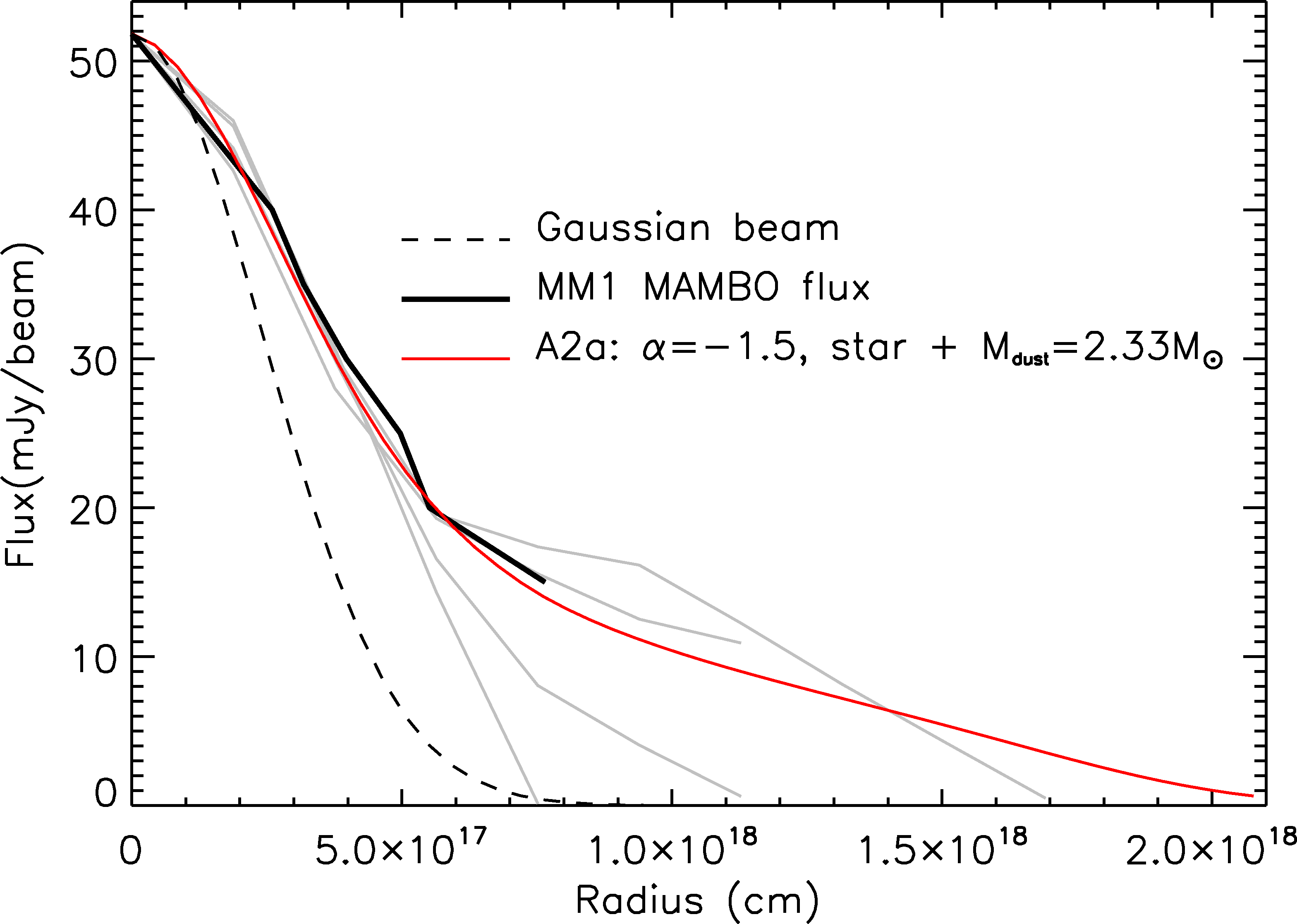}
                %\caption{Models 1}
                %\label{fig:models1}
        \end{subfigure}\\
        \begin{subfigure}[t]{0.56\textwidth}
                \includegraphics[width=1\textwidth]{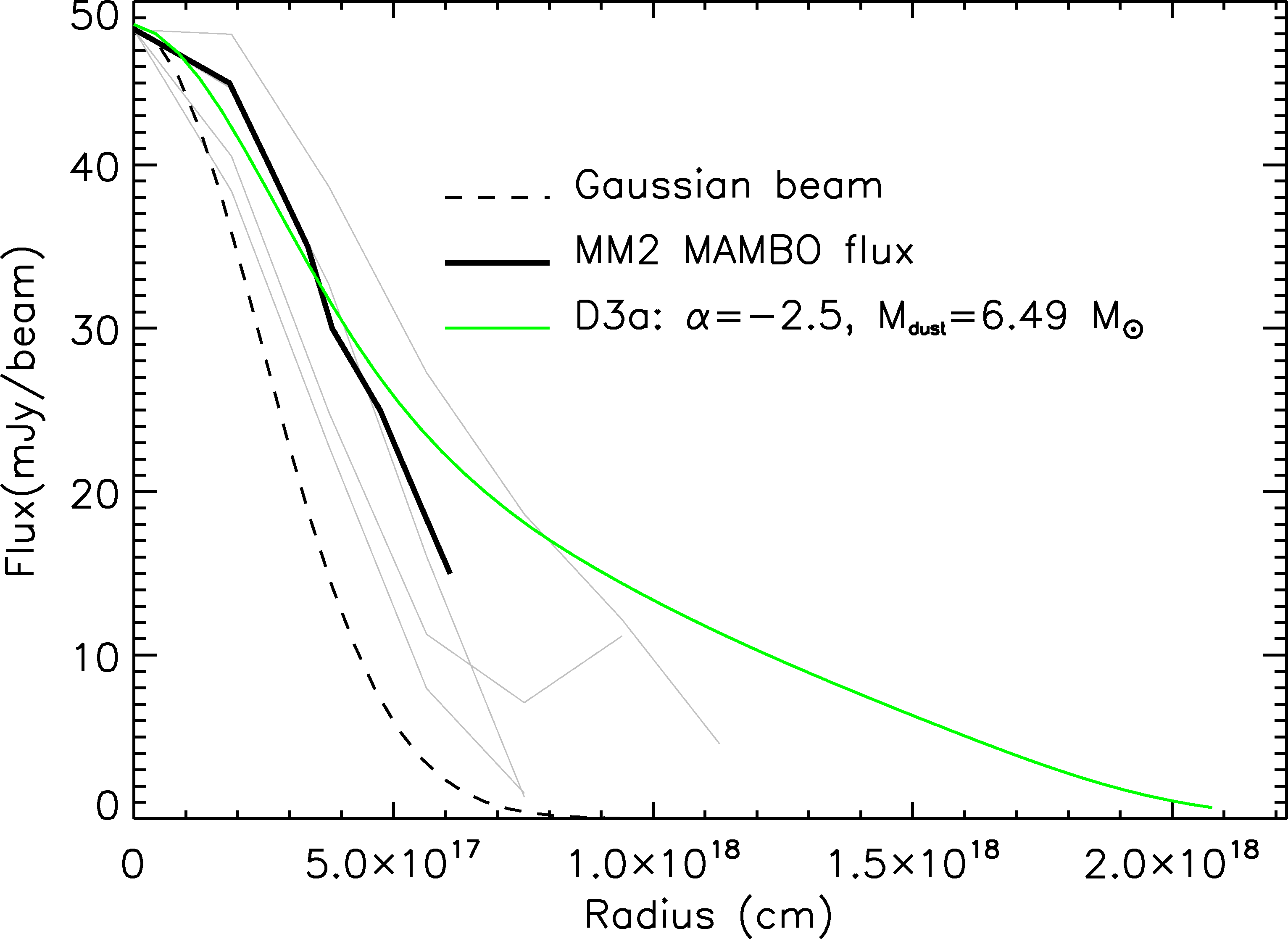}
                %\caption{Models 1}
                %\label{fig:models1}
        \end{subfigure}\\
        \begin{subfigure}[t]{0.58\textwidth}
                \includegraphics[width=1\textwidth]{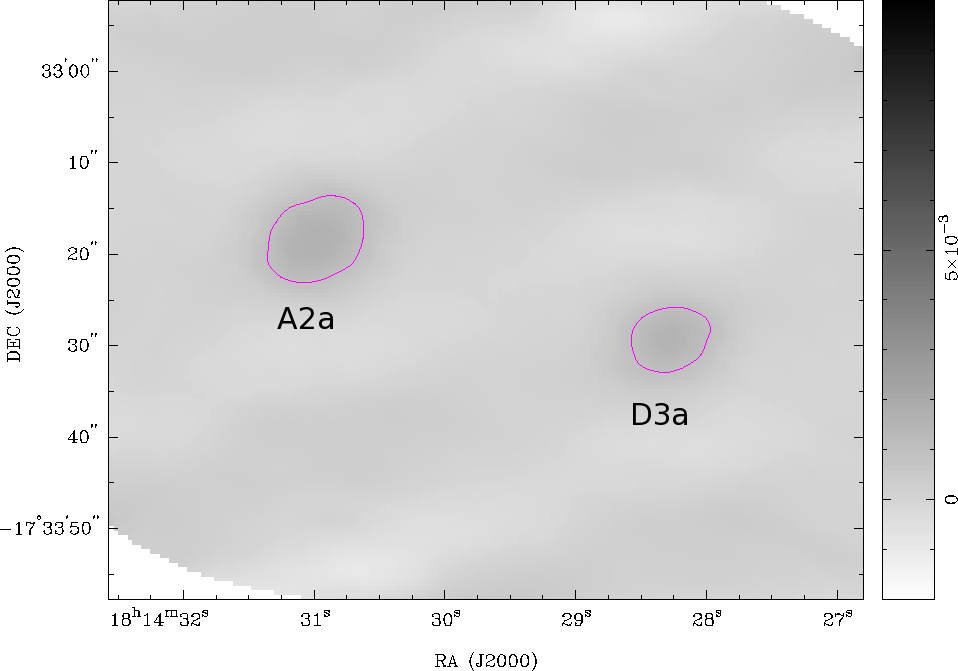}
                %\caption{Models 2}
                %\label{fig:models2}
        \end{subfigure}
\caption{CASA simulated SMA 1.3mm continuum map of the SDC13 region (bottom) made using RADMC-3D models A2a (top) and D3a (centre; See Table \ref{tab:sim_models}). The map contour is at 1.5mJy. The rms noise in the centre of the map $\rm{\sim0.09mJy}$. Model fluxes for A2a (shown in red) and D3a (shown in green) are compared with plots of the observed MAMBO 1.2mm flux against equivalent radius for MM1 and MM2 respectively (solid black lines). Model parameters are given in the key of each plot. The grey lines in the top two plots indicate perpendicular slices of observed flux against radius; these slices give an indication of the range of variation in the observed flux distributions in MM1 and MM2. The dashed curves indicate the MAMBO beam (FWHM=$10.7^{\prime\prime}$). }
\label{fig_model_map_original}
\end{figure*}

\begin{figure*}
	\centering
        \begin{subfigure}[t]{0.56\textwidth}
                \includegraphics[width=1\textwidth]{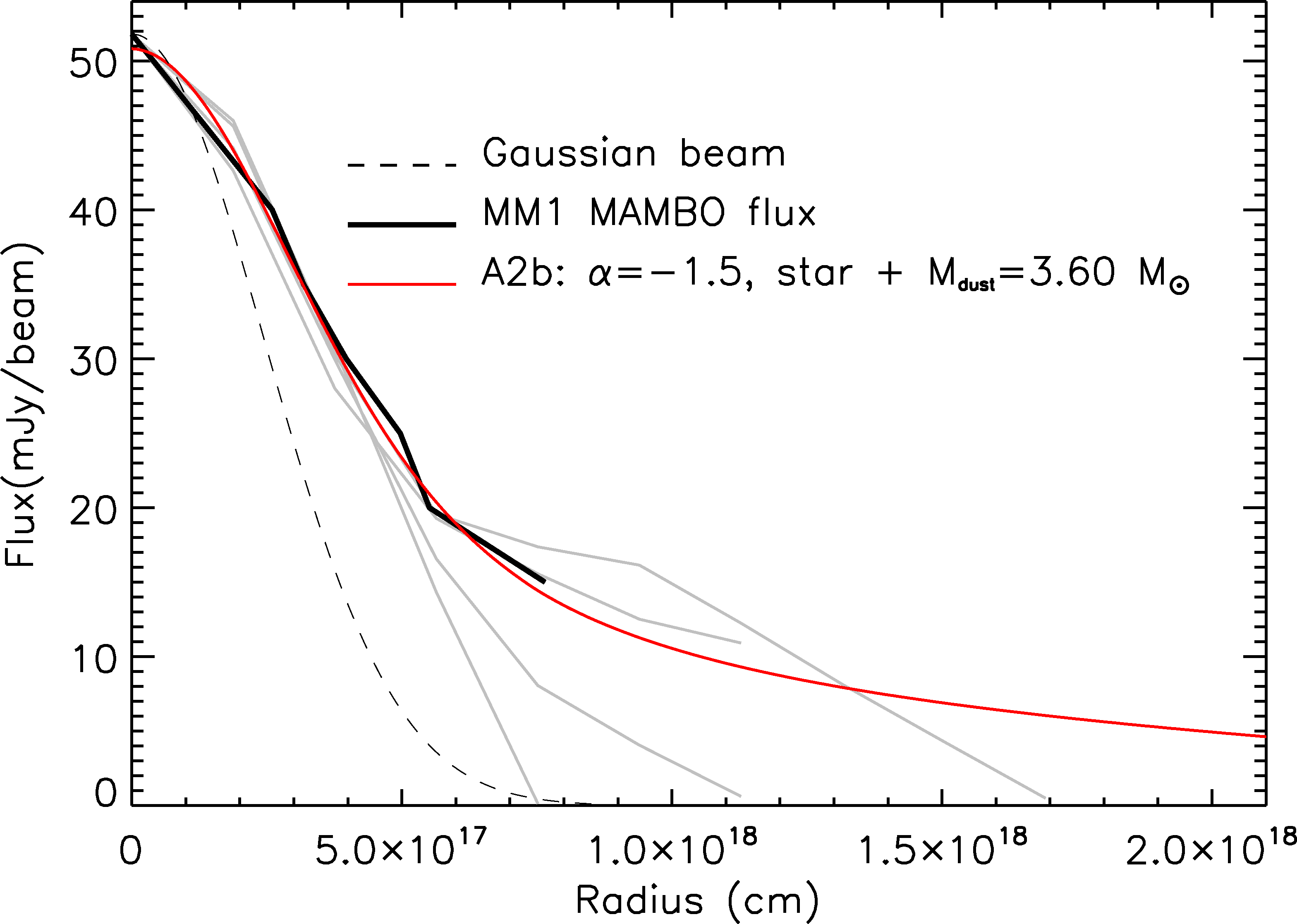}
                %\caption{Models 1}
                %\label{fig:models1}
        \end{subfigure}\\
        \begin{subfigure}[t]{0.55\textwidth}
                \includegraphics[width=1\textwidth]{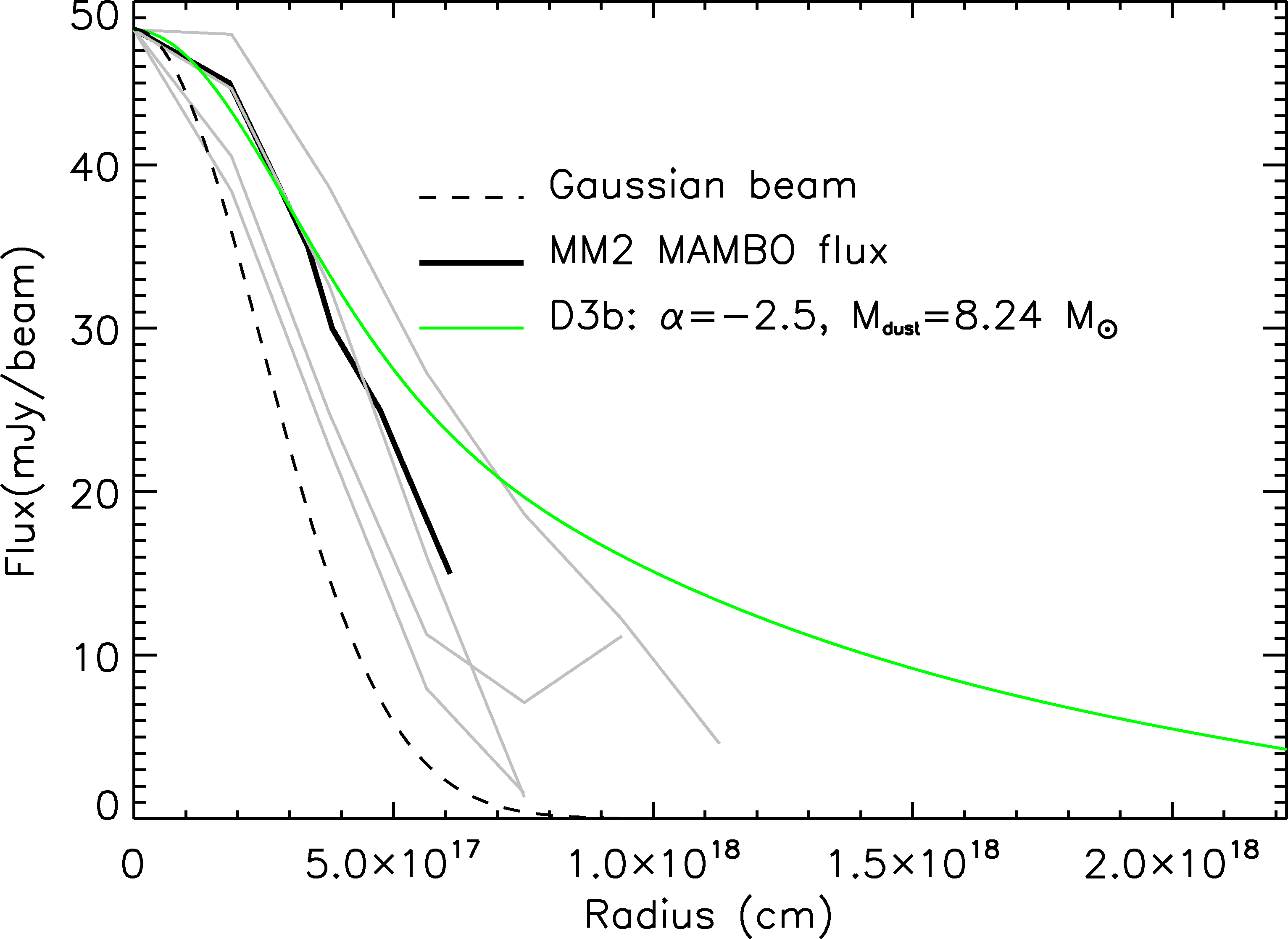}
                %\caption{Models 1}
                %\label{fig:models1}
        \end{subfigure}\\
        \begin{subfigure}[t]{0.57\textwidth}
                \includegraphics[width=1\textwidth]{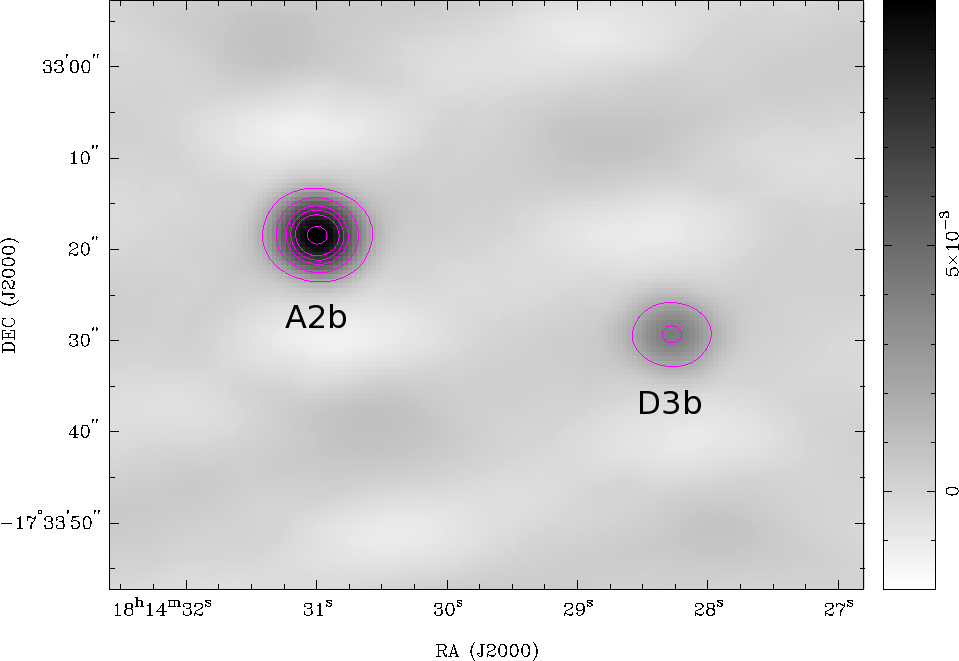}
                %\caption{Models 2}
                %\label{fig:models2}
        \end{subfigure}
\caption{CASA simulated SMA 1.3mm continuum map of the SDC13 region (bottom) made using RADMC-3D models A2b (top) and D3b (centre; See Table \ref{tab:sim_models}). Map contours are at 2, 4, 6, 8 and 10 mJy. The rms noise in the centre of the map $\rm{\sim0.3mJy}$. Model fluxes for A2b (shown in red) and D3b (shown in green) are compared with plots of the observed MAMBO 1.2mm flux against equivalent radius for MM1 and MM2 respectively (solid black lines). Model parameters are given in the key of each plot, with further details given in the text. The grey lines in the top two plots indicate perpendicular slices of observed flux against radius; these slices give an indication of variation in the observed flux distributions in MM1 and MM2. The dashed curves indicate the MAMBO beam (FWHM=$10.7^{\prime\prime}$). }
\label{fig_model_map_adjusted}
\end{figure*}

Finally, imaging of the visibility data was carried out using the CASA \texttt{clean} task in mosaic mode, with a clean box at each source position and a threshold of $\sim$0.45mJy, (a mid point value between the $\sim$0.3mJy for the compact data, and $\sim$0.6mJy for the extended data). The cell size ($0.52^{\prime\prime}\times0.50^{\prime\prime}$) and the restoring beam ($3.78^{\prime\prime}\times2.52^{\prime\prime}$, PA$=41.55^{\circ}$) were fixed to match the real data image values. Examples of the results of the CASA simulation process are shown in Figures \ref{fig_model_map_original} and \ref{fig_model_map_adjusted}.

Figure \ref{fig_model_map_original} shows the result of the CASA simulation using models A2a and D3a (see Section \ref{sec:modelling} and Table \ref{tab:sim_models}). The resulting sub-fragments A2a and D3a are of similar size and flux, with peak fluxes $\rm{F_{A2a}}=1.68$ mJy/beam and $\rm{F_{D3a}}=1.52$ mJy/beam respectively (i.e $\rm{F_{A2a}\sim10\%}$ larger than $\rm{F_{D3a}}$). In comparison, the peak fluxes of sub-fragments A and D in the SMA 1.3mm data (coincident with MM1 and MM2 in the MAMBO data) are $\rm{F_{A}}=13.6$ mJy/beam and $\rm{F_{D}}=4.9$ mJy/beam respectively i.e $\rm{F_{A}}$ $\sim{2.8}$ times larger than $\rm{F_{D}}$ (see Table \ref{tab:source_properties}). We find we were able to better match the observed peak fluxes F$_{A}$ and F$_{D}$ (and therefore their ratio $\rm{F_{A}}/\rm{F_{D}}$) by adjusting the value of $\rm{R_{out}}$ in the RADMC-3D models of MM1 and MM2.

We normalize our RADMC-3D models based on the total dust mass $\rm{M_{dust}}$, which we use to calculate a normalisation constant $\rho_{0}$ (the density at the outer radius $\rm{R_{out}}$). For a density distribution ${\rho}=cR^{\alpha}$ and ${\rho}_{0}=cR_{out}^{\alpha}$. The density distribution can thus be expressed as ${\rho}={\rho}_{0}(R/R_{out})^{\alpha}$. Based on the dust mass 
\begin{equation}\label{eq:dust_mass}
M_{dust}=\int_{R_{in}}^{R_{out}} \! {4{\pi}R^{2}\rho} \, \mathrm{d}R=\int_{R_{in}}^{R_{out}} \! {4{\pi}R^{2}{\rho}_{0}R_{out}^{-\alpha}R^{\alpha}} \, \mathrm{d}R, 
\end{equation}
we find the normalisation constant
\begin{equation}\label{eq:norm_constant}
\rho_{0}=\frac{M_{dust}(3+\alpha)}{4{\pi}R_{out}^{-\alpha}(R_{out}^{3+\alpha}-R_{in}^{3+\alpha})}.
\end{equation}
A consequence of this method of normalisation is that $\rm{M_{dust}}\propto\rm{R_{out}}$ i.e. an increase/decrease in $\rm{R_{out}}$ in our models requires an increase/decrease in $\rm{M_{dust}}$ to maintain the same density distribution. This increase/decrease in $\rm{M_{dust}}$ results in an increase/decrease in flux from the model. 

For example, as shown in Figure \ref{fig_model_map_adjusted}, we are better able to match the observed fluxes $\rm{F_{A}}$ and $\rm{F_{D}}$ with increased outer radii for the RADMC-3D models of MM1 and MM2 (Model A2b with $\rm{R_{A2b}=3.5\times10^{18}cm}$ and Model D3b with $\rm{R_{D3b}=2.9\times10^{18}cm}$ respectively; see Table \ref{tab:sim_models}). This subsequently increases the model peak fluxes, giving values $\rm{F_{A2b}=10.5mJy}$ and $\rm{F_{D3b}=4.2mJy}$ resulting in a flux ratio $\rm{F_{A2b}/F_{D3b}\sim2.5}$. This is similar to the ratio of the observed fluxes for sub-fragments A and D, $\rm{F_{A}/F_{D}}\sim{2.8}$. Figure \ref{fig_model_map_adjusted} indicates that adjusting the model outer radii as described above does not have a significant effect on the quality of the fit to the MAMBO observations.

Simulations performed using models of MM2 with truncated and two-part power-law density profiles produce model sub-fragments that are too diffuse to be detected. As such, we consider models A2b and D3b to be the \emph{best-fits} for MM1 and MM2 respectively, as they provide the closest match to both the MAMBO and SMA observations. The best fit model to MM1 is therefore a cloud of total mass $\sim360\rm{M_{\odot}}$ (assuming gas:dust=100:1), radius $\rm{R_{out}=R_{A2b}=3.5\times10^{18}cm}$ and density profile $\rm{r\propto\rho^{-1.5}}$, with a central star of temperature $\rm{T_{\star}=14914K}$. The best fit model to MM2 has a total mass $\sim824\rm{M_{\odot}}$, radius $\rm{R_{out}=R_{D3b}=2.9\times10^{18}cm}$ and density profile $\rm{r}\propto\rho^{-2.5}$ and does not contain a central star. \citet{Peretto2014} find masses of 74.8$\rm{M_{\odot}}$ in 0.26 pc and 81.1$\rm{M_{\odot}}$ in 0.21 pc for MM1 and MM2 respectively (see Table \ref{tab:source_properties}). Within the same radii, our best-fit models for MM1 and MM2 contain masses 41.3$\rm{M_{\odot}}$ and 254$\rm{M_{\odot}}$ respectively.

%(Table \ref{tab:twopart_bestfit})
\renewcommand{\tabcolsep}{14pt}
\begin{table}
\caption{Parameters of RADMC-3D models shown in Figures \ref{fig_model_map_original} and \ref{fig_model_map_adjusted} and their corresponding CASA simulated peak flux values.}
% title of Table
%\small{
\centering
% used for centering table
% is used to refer this table in the text
\begin{tabular}{l l l l l}
% centered columns (4 columns)
\hline\hline
% inserts double horizontal lines
						
Model	&$\alpha$	&$R_{out}$	&$M_{dust}$	&$\rm{F_{peak}}$	\\
%\cline{8-9}
% table heading
	&		&($10^{18}$cm)	&(M$_{\odot}$)	&(mJy)			\\
%units
\hline
% inserts single horizontal line
A2a	&-1.5		&1.9		&2.3		&1.7			\\
D3a	&-2.5		&1.9		&6.5		&1.5			\\
A2b	&-1.5		&3.5		&3.6		&10.5			\\
D3b	&-2.5		&2.9		&8.2		&4.2			\\
%body of the table
\hline
%inserts single line
\end{tabular}
\tablefoot{$\rm{F_{peak}}$ is the CASA simulated peak flux, $\alpha$ is the index of the density profile of the model core and $\rm{R_{out}}$ is the outer radius of the model core. M$_{\rm{dust}}$ is the best-fit model dust mass.}
\label{tab:sim_models}
\end{table}

%\documentclass[12pt,a4paper,final]{report}%size 10 or 12 depending on font used
%\usepackage{setspace}%must be 1.5 or double spaced except capts etc-setspace does this
%\usepackage[left=40mm,right=25mm,top=25mm,bottom=25mm]{geometry}%min left 40mm rest 15mm
%\usepackage{amssymb}
%\usepackage{natbib}
%\usepackage{aas_macros}
%\usepackage{graphicx}
%\usepackage{caption}
%\usepackage{subcaption}
%\usepackage{subfig}
%\usepackage{sidecap}
%\usepackage{float}
%\usepackage{rotating}
%\usepackage{array}
%\usepackage{tabularx}
%\usepackage{multirow}
%\usepackage{subeqn}
%\usepackage{pifont}
%\bibliographystyle{apj3} 
%\doublespacing
%\begin{document}
%\end{document}

%\documentclass[12pt,a4paper,final]{report}%size 10 or 12 depending on font used
%\usepackage{setspace}%must be 1.5 or double spaced except capts etc-setspace does this
%\usepackage[left=40mm,right=25mm,top=25mm,bottom=25mm]{geometry}%min left 40mm rest 15mm
%\usepackage{amssymb}
%\usepackage{natbib}
%\usepackage{aas_macros}
%\usepackage{graphicx}
%\usepackage{caption}
%\usepackage{subcaption}
%\usepackage{subfig}
%\usepackage{sidecap}
%\usepackage{float}
%\usepackage{rotating}
%\usepackage{array}
%\usepackage{tabularx}
%\usepackage{multirow}
%\usepackage{subeqn}
%\usepackage{pifont}
%\bibliographystyle{apj3} 
%\doublespacing
%\begin{document}
%\end{document}

\section{Discussion and Summary}
High resolution SMA 1.3mm observations toward MM1 and MM2, the two largest fragments (i.e. cores) in SDC13 (as observed with MAMBO at 1.2mm with lower resolution), indicate the presence of 4 sub-fragments A, B, C and D (Figure \ref{fig:continuum}). Three of the sub-fragments A, C and D are associated with the cloud. One of the sub-fragments, B, does not appear to be associated with a significant extended envelope and is much less prominent in the MAMBO 1.2mm observations (Figure \ref{fig:mambo_contours}). The nature of B remains unconfirmed, but statistical analysis indicates that B is unlikely to be a background source, an AGB star or free-free emission of a HII region. However B could plausibly be a runaway object (see Section \ref{sec:fragmentB}).

MM1 consists of two sub-fragments: A, at its centre and a fainter sub-fragment C. The fragmentation of MM1 into two objects is confirmed by \textit{Herschel} Hi-GAL observations at 70$\mu{m}$ (see Figure \ref{fig:herschel}). MM2 consists of only one object at its centre (sub-fragment D). Although MM1 and MM2 look similar at the lower MAMBO resolution, with a similar size and brightness, at the higher resolution of the SMA, they look very different (Figure \ref{fig:mambo_contours}), with the object at the centre of MM1 (sub-fragment A) appearing much brighter than the object at the centre of MM2 (sub-fragment D). We are able to reproduce this effect with CASA simulations of SMA observations, using RADMC-3D models with single power law density distributions and extended outer radii (see Figure \ref{fig_model_map_adjusted}, Section \ref{sec:simulation}). 

MM1 is associated with 8$\rm{\mu{m}}$ emission indicative of active star formation. MM2 shows no evidence of emission at 8$\mu{m}$ or $24\rm{\mu{m}}$ making it a good candidate for a prestellar core (Figure \ref{fig:mambo_spitzer}). MM2 requires a steeper density profile and higher mass to model its emission ($\rm{r\propto\rho^{-2.5}}$; $\rm{M_{dust}}$$\sim$$824\rm{M_{\odot}}$) compared to those required for MM1 ($\rm{r\propto\rho^{-1.5}}$; $\rm{M_{dust}}$$\sim$$360\rm{M_{\odot}}$). The relatively steep denisty profile required to model MM2 depends on a significant temperature decrease at its centre (see Section \ref{sec:modelling}), which can be justified by the lack of star formation in the core. Increasing the minimum temperature decreases the required model mass, but for temperatures $\sim{10}$ K, MM2 still requires a steeper model density profile than MM1.

A further example of a higher-mass core at an earlier evolutionary stage than a neighbouring lower-mass core, can be found in \cite{Stephens2015}.

\subsection{The Future Stellar Population in SDC13}\label{sec:imf}

Based on our models, MM1 contains one star with a  mass of $\sim7.36\rm{M_{\odot}}$. We can use this to estimate the possible future stellar population which MM1 could produce, by assuming that the stars will have an IMF distribution of masses, and conservatively normalising the IMF to have one star in the mass range $6\rm{M}_{\odot}\le\rm{M}_{\star}\le9{M}_{\odot}$. Adopting a power law form of the IMF, $\rm{dN}\propto\rm{M}^{{-}\alpha}\rm{dM}$ \citep[where $2.3\le\alpha\le{2.35}$,][]{Salpeter1955,Kroupa2002}, we find that MM1 could potentially form $\sim25$--$26$ stars in the mass range $1\rm{M}_{\odot}\le\rm{M}_{\star}\le120{M}_{\odot}$. \citet{Kroupa2002} estimate that stars in this mass range make up $\sim$6.1$\%$ of the total number of stars in the galactic field (for an IMF with $\alpha={2.3}$). This suggests that MM1 has the potential to form at least $\sim$409 stars in total (with masses $0.01\rm{M}_{\odot}\le\rm{M}_{\star}\le120{M}_{\odot}$). Based on the average stellar mass ($\rm{\bar{m}}$=0.38$\rm{M}_{\odot}$) given by \citet{Kroupa2002}, this equates to a  total stellar mass $\sim$155$\rm{M}_{\odot}$. Using the total mass of MM1 ($\sim$360$\rm{M}_{\odot}$, derived from the best-fit RADMC-3D model; Section \ref{sec:simulation}), this is equivalent to a star formation efficiency of $\sim$43$\%$, consistent with the relatively high star formation efficiencies seen in regions of dense core and cluster formation \citep[e.g.][]{Wilking1983, Hillenbrand1998, Koenig2008}. MM2 might be expected to form a similar population of stars to MM1 in the future.
 
\subsection{The Evolution of MM1 and MM2}

The absence of a star in MM2 suggests that it is in an earlier stage of evolution than MM1. However, based on the proximity of MM1 and MM2, one could assume that they were formed at the same time. This would suggest a shorter evolution timescale for MM1 than for MM2. Differences in star formation timescales could be the result of differences in the size and density of the region in a large scale dynamical collapse scenario \citep[e.g.][]{Peretto2014} or differences in turbulent and/or magnetic support between regions \citep{Myers1993, McKee2003}. \citet{Peretto2014} find a slightly higher value for the velocity dispersion in the vicinity of MM2 ($\rm{\sigma_{tot}}\sim$0.84 kms$^{-1}$) than in the vicinity of MM1 ($\rm{\sigma_{tot}}\sim$0.81 kms$^{-1}$), however the difference is small and does not translate into enough additional support to prevent the collapse of MM2. The slight difference in the measured velocity dispersion of the turbulent motions around MM1 and MM2 is therefore not a plausible explanation for their different evolutionary stages.

In order to explain the observed distribution of protostellar and prestellar cores in the Perseus molecular cloud, \citet{Hatchell2008} suggest that higher-mass prestellar cores will have shorter evolutionary timescales. If this is the case, MM1 may have had a larger initial mass than MM2. Based on our best fit models however, MM2 is currently $\sim460$M$_{\odot}$ more massive than MM1 (Section \ref{sec:simulation}). It is possible that MM1 could have experienced mass loss after the formation of its central protostar via accretion on to the star, but, based on our best fit model, only a small proportion of the mass difference can be accounted for in the star at the centre of MM1 ($\sim7$ M$_{\odot}$). 

A further possibility is that the observed mass difference between MM1 and MM2 is the result of outflows from the stars in MM1. The three-colour image of SDC13 shown in Figure \ref{fig:mambo_spitzer} indicates an extended green (8.0$\mu$m) biconical structure in the vicinity of MM1, characteristic of emission from PAHs excited by UV photons. IR emission in the Spitzer IRAC bands, observed towards regions of star formation, has been shown to trace the illumination of outflow cavities by scattered light from central star forming objects \citep{Keping2006, Tobin2007, Velusamy2011}. The presence of 8$\mu$m biconical emission around MM1 provides evidence for the existence of a bipolar outflow cavity (and therefore a bipolar outflow) in this region. However, the removal of $\gtrsim{460}$ M$_{\odot}$ would imply an unusually high mass loss rate $\sim5\times10^{-4}$--$5\times10^{-3}$ M$_{\odot}$yr$^{-1}$ over $10^{5}$--$10^{6}$ years, compared to a typical value of $\sim10^{-4}$ M$_{\odot}$yr$^{-1}$ for similarly sized clumps \citep[e.g.][]{Zhang2005,deVilliers2014}.

An alternative explanation for the mass difference between MM1 and MM2 could be that MM2 may have continued to increase in mass once MM1 started star formation. However, using the mass inflow rate estimated by \citet{Peretto2014} ($\sim{2.5\times10^{-5}}$ M$_{\odot}$yr$^{-1}$) this would imply an age difference $\sim10^{7}$ years between MM1 and MM2. This is unrealistically long given the presence of embedded young stars in MM1 and the $\sim10^5$\,yr timescale for the formation of massive stars \citep[e.g.][]{McKee2002}.

The density profile in MM1 is consistent with that expected for a collapsing region whereas the profile of MM2 is steeper, or at least has a steeper outer envelope. This difference may reflect how a clump density profile evolves as the clump evolves towards forming stars. With its apparent coeval, similar mass, but star forming, neighbouring clump in a near identical environment, MM2 is an important laboratory for further study towards understanding how massive clumps evolve towards the formation of massive stars. The steep density profile of MM2 may also make it more likely to form a single massive central object. This is based on evidence presented by \citet{Girichidis2011}, who find that massive stars predominantly form from cores with initial density profiles that are strongly centrally condensed (independent of  support from radiation or magnetic fields). They conclude that the initial density profile of star forming cores is perhaps the most important factor in determining the fragmentation, evolution and final mass distribution of massive star forming regions.

\bibliography{AA-2015-27062} %Y
\end{document}